\documentclass[prd, twocolumn, nofootinbib]{revtex4}
\usepackage{graphicx}
\usepackage{dcolumn}
\usepackage{epsfig} 
\usepackage{amsmath}
\usepackage{amsfonts}
\usepackage{graphicx}
\usepackage{amssymb}
\usepackage{color}

\begin{document}
\title{The Imprint of f(R) Gravity on Non-Linear Structure Formation}

\author{Ixandra Achitouv$^{1,2}$\footnote{E-mail:iachitouv@swin.edu.au}}

\author{Marco Baldi$^{3,4,5}$}

\author{Ewald Puchwein$^{6}$}

\author{Jochen Weller$^{7,8,9}$}

\address{ 
$^{1}$Centre for Astrophysics \& Supercomputing, Swinburne University of Technology, P.O. Box 218, Hawthorn, VIC 3122, Australia \\
$^{2}$ARC Centre of Excellence for All-sky Astrophysics (CAASTRO), 44 Rosehill St, Redfern, NSW 2016, Australia\\  
$^{3}$Dipartimento di Fisica e Astronomia, Alma Mater Studiorum Universit\`a di Bologna, viale Berti Pichat, 6/2, I-40127 Bologna, Italy\\
$^{4}$INAF - Osservatorio Astronomico di Bologna, via Ranzani 1, I-40127 Bologna, Italy\\
$^{5}$INFN - Sezione di Bologna, viale Berti Pichat 6/2, I-40127 Bologna, Italy\\
$^6$ Kavli Institute for Cosmology Cambridge and Institute of Astronomy, University of Cambridge, Madingley Road, Cambridge CB3 0HA, UK\\
$^7$ Universit\"ats-Sternwarte, Fakult\"at f\"ur Physik, Ludwig-Maximilians Universit\"at M\"unchen, Scheinerstr. 1, 81679 M\"unchen, Germany\\
$^8$Excellence Cluster Universe, Boltzmannstr. 2, 85748 Garching,  Germany\\
$^9$Max Planck Institute for Extraterrestrial Physics, Giessenbachstr. 1, 85748 Garching, Germany
}

\begin{abstract}
We test the imprint of $f(R)$ modified gravity on the halo mass function, using N-body simulations and a theoretical model developed in \citep{Koppetal}. We find a good agreement between theory and simulations $\sim 5\%$. We extend the theoretical model to the conditional mass function and apply it to the prediction of the linear halo bias in $f(R)$ gravity. Using the halo model we obtain a prediction for the non-linear matter power spectrum accurate to $ \sim 10\%$ at $z=0$ and up to $k=2{\rm h}/{\rm Mpc}$. We also study halo profiles for the $f(R)$ models and find a deviation from the standard general relativity result up to $40 \%$, depending on the halo masses and redshift. This has not been pointed out in previous analysis. Finally we study the number density and profiles of voids identified in these $f(R)$ N-body simulations. We underline the effect of the bias and the sampling to identify voids. We find significant deviation from GR when measuring the $f(R)$ void profiles with $f_{\rm R0} < -10^{-6}$.  

\end{abstract} 
\maketitle

\section{Introduction}

As cosmology is approaching the epoch of precision measurements with a number of current and upcoming observational enterprises -- as e.g. BOSS \citep{Ahn_etal_2013}, HETDEX \citep{HETDEX}, DES \citep{DES}, DESI \cite{DESI}, eBoSS \cite{eBOSS}, WFIRST \cite{WFIRST},
LSST \citep{LSST} and EUCLID \citep{EUCLID-r} -- aiming to constrain the basic cosmological parameters to percent precision, significant efforts are required to improve theoretical and numerical modeling of possible deviations from the standard $\Lambda $CDM scenario. In particular, some of the most ambitious observational enterprises of the next decade (such as e.g. DES and Euclid) will focus on testing and constraining possible alternatives to the cosmological constant $\Lambda $ as the source of the observed accelerated expansion of the Universe \citep[][]{Riess_etal_1998,Perlmutter_etal_1999}. In fact, despite the cosmological constant remaining so far the simplest possible explanation of the cosmic acceleration that consistently accounts for available observations \citep[see e.g.][]{Blake_etal_2011,Planck_016, eBOSS}, its theoretical footings represent an open problem for both cosmology and theoretical particle physics as the observed energy scale associated with $\Lambda $ appears to be unnaturally small (e.g. \cite{Bousso}). 

A possible extension to the cosmological constant paradigm are represented by a dynamical Dark Energy field whose background density  evolves in time along the cosmic history \citep[see e.g.][]{Wetterich_1988,Ratra_Peebles_1988} -- possibly also featuring direct interactions with other matter fields \citep{Wetterich_1995,Amendola_2000,Farrar2004,Amendola_Baldi_Wetterich_2008,Baldi_2011a}. Another possibility are modifications of General Relativity (GR) on large scales \citep[as e.g. in][]{Buchdahl_1970,Starobinsky_1980,Hu_Sawicki_2007,Sotiriou_Faraoni_2010,Dvali_Gabadadze_Porrati_2000,Nicolis_Rattazzi_Trincherini_2009}, resulting in a modification of the gravitational instability processes that leads to the formation of cosmic structures. The simplest and most widely investigated modification of the gravitational theory with a relevant impact on cosmology is the class of scalar-tensor theories known as $f(R)$ gravity, where the standard Einstein-Hilbert Action is modified by adding to the Ricci scalar $R$ a further term given by an arbitrary function $f(R)$:
\begin{equation}
\label{fRaction}
  S = \int {\rm d}^4x \, \sqrt{-g} \left( \frac{R+f(R)}{16 \pi G} + {\cal L}_m \right),
\end{equation}
where $G$ is Newton's gravitational constant,  $g$ is the determinant of the metric
tensor $g_{\mu \nu }$, and ${\cal L}_m$ is the Lagrangian of matter content of the Universe.
By suitably choosing the form of the function $f(R)$ it is possible to tune the model so to have an expansion history arbitrarily close to the standard $\Lambda $ cold dark matter (CDM) scenario \citep{Hu_Sawicki_2007}, thereby matching present background observations and confining possible deviations from the standard model only in the evolution of density perturbations in both the linear and nonlinear regimes \citep[][]{Pogosian_Silvestri_2008,Oyaizu_etal_2008,Li_etal_2012,Puchweinetal,Llinares_Mota_Winther_2013}. 

Consistency with solar system tests of scalar-tensor theories \citep{Bertotti_etal_2003, Will_2005} additionally requires that standard General Relativity is recovered in our local environment, which can be ensured in $f(R)$ theories by the so-called {\em Chameleon} screening mechanism \citep{Khoury_Weltman_2004} that suppresses the deviation from standard gravity in high-density environments such as the Milky Way.\\

In this work, we use the N-body simulations described in \citep{Puchweinetal} to look at the imprint of $f(R)$ models on the abundance of halos. We compare the results of the N-body simulations to the results of the spherical collapse model developed in \citep{Koppetal}. We extend this theoretical investigation to the conditional mass function with a drifting diffusive barrier and a mean value, which depends on the halo masses according to the $f(R)$ models we consider. We deduce the linear halo bias for $f(R)$ gravity, which allows us to test the prediction of the non-linear matter power spectrum using the halo model \citep{CooraySheth}. We also compare the halo profiles we measure in the simulations to the GR case. Finally, we study the imprint of $f(R)$ models on the void profiles and void abundances. These are particularly interesting for considering deviations from GR. We measure void profiles in two cases: firstly, we identify the voids in each simulation and stack their profiles. Secondly, we identify the voids in the GR simulation which have the same initial conditions as the $f(R)$ simulations and measure the profiles in the $f(R)$ simulation by using the GR positions of the voids. 

 The outline of this paper is as follows: In section \ref{secMF} we present the N-body simulation results for the mass function in $f(R)$ and compare them to the model presented in \cite{Koppetal}. In section \ref{secCondi} we predict the linear halo bias for $f(R)$ models. In section \ref{secPro} we study the halo profiles in $f(R)$ simulations and the deviation compared to the GR results. In section \ref{secHM} we use the halo model to predict the non-linear power spectrum in the $f(R)$ and compare to the GR results. In Sec.\ref{secVA} we measure the abundance of voids, while in section \ref{secVP} we study the profile of voids in the $f(R)$ simulations. Finally, in section \ref{secConclu} we draw our conclusions.

\section{Mass function}\label{secMF}
\begin{figure}[ht]
\centering
\begin{tabular}{cc}
\includegraphics[scale=0.35]{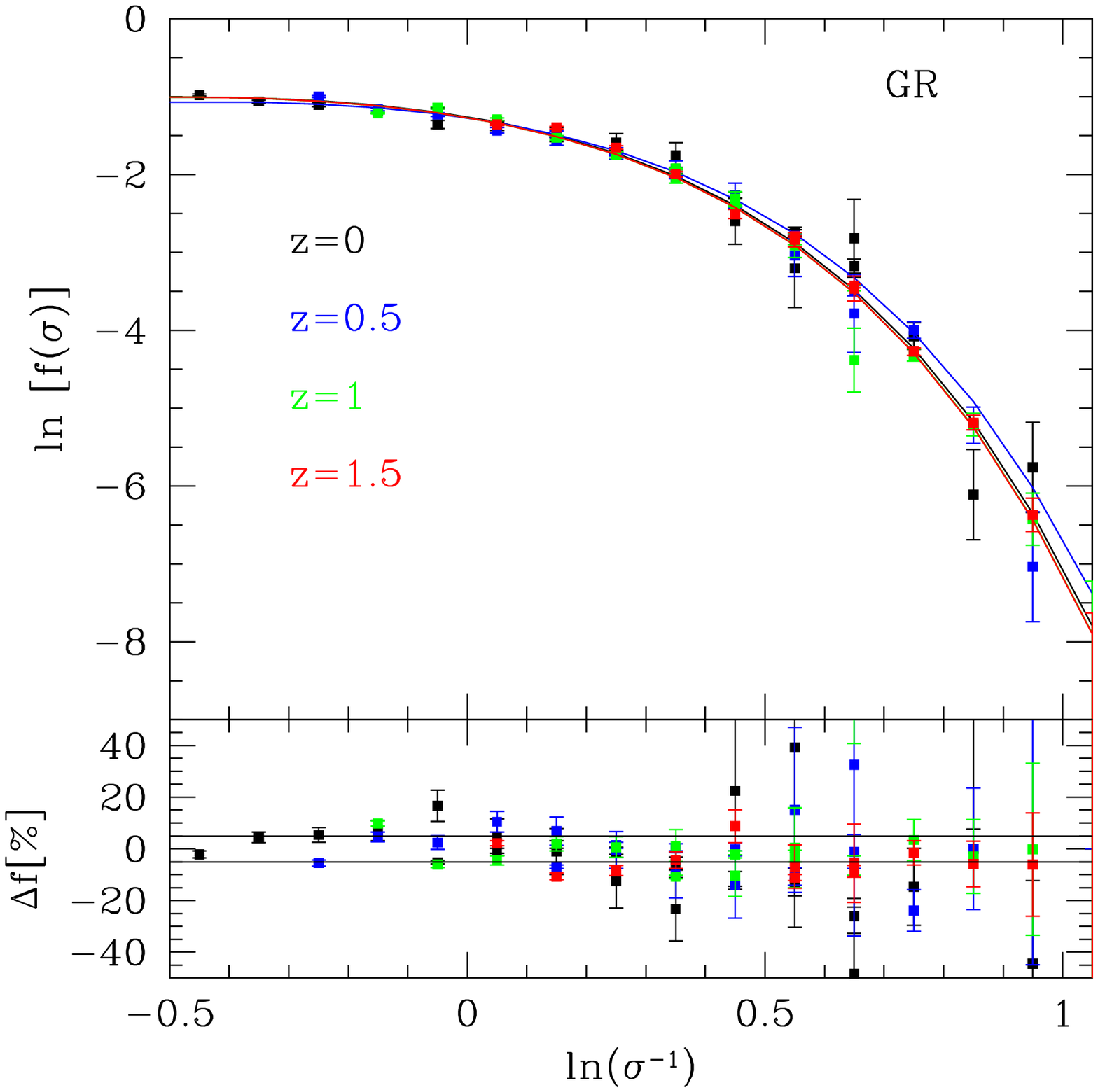}\\
\includegraphics[scale=0.35]{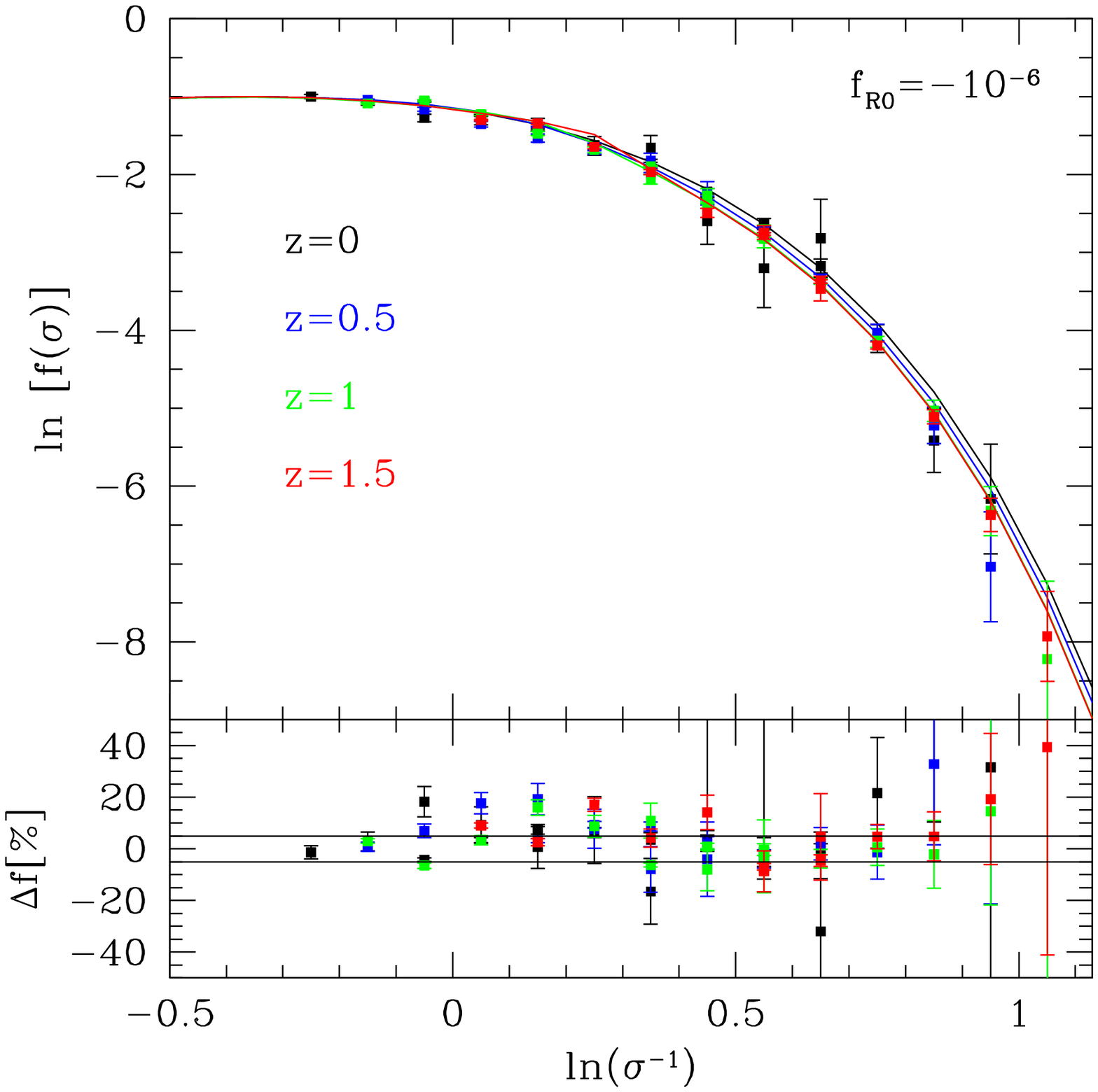}
\end{tabular}
\caption{Multiplicity function for GR ($\Lambda$CDM) and $f_{\rm R0}=1-0^{-6}$ at various redshifts. Solid lines are theory, squares are from simulations. Lower panels show the relative difference. Black solid lines show $\pm5\%$ differences.}\label{Fig1}
\end{figure}
\begin{figure}[ht]
\centering
\begin{tabular}{cc}
\includegraphics[scale=0.35]{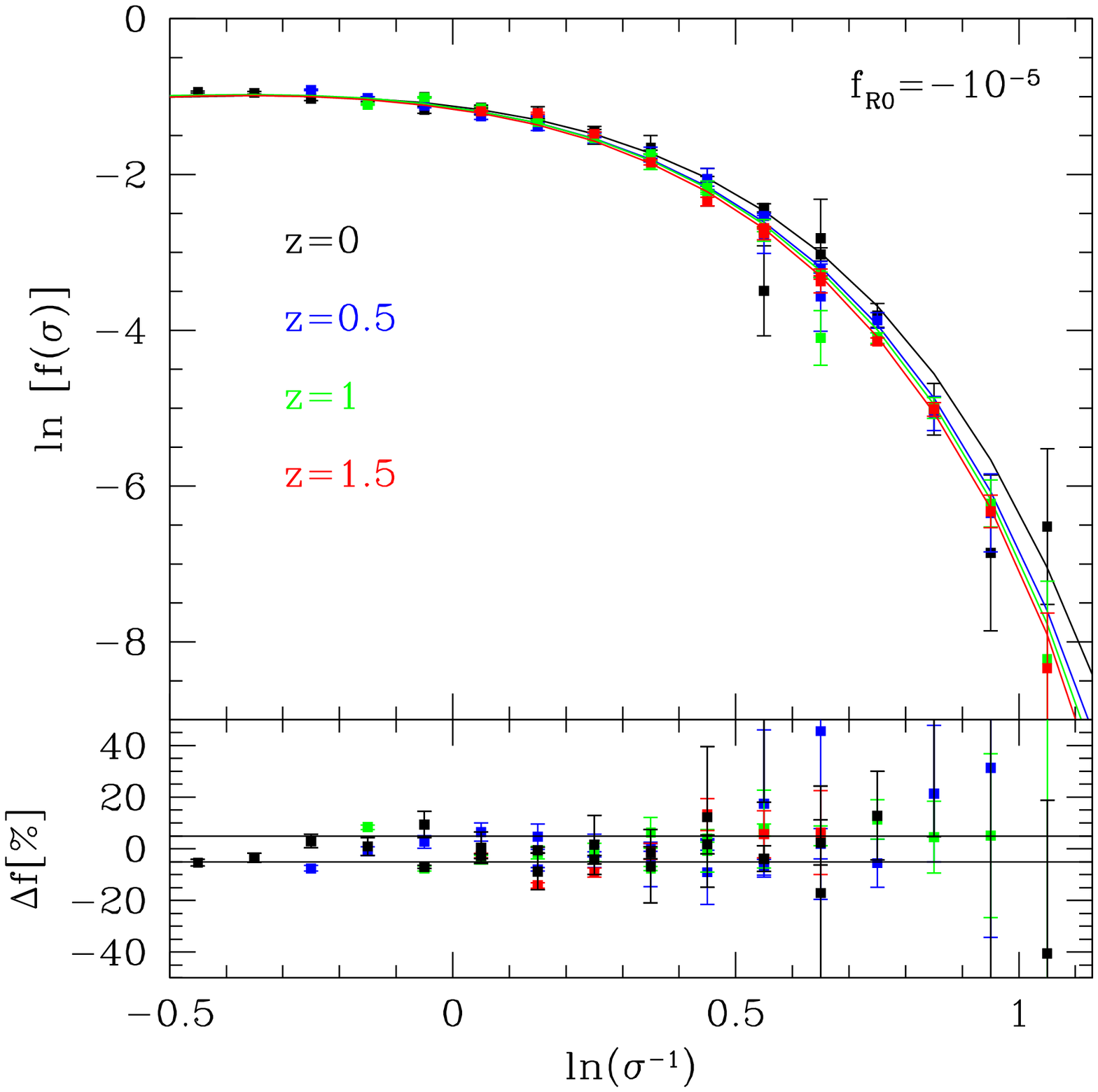}\\
\includegraphics[scale=0.35]{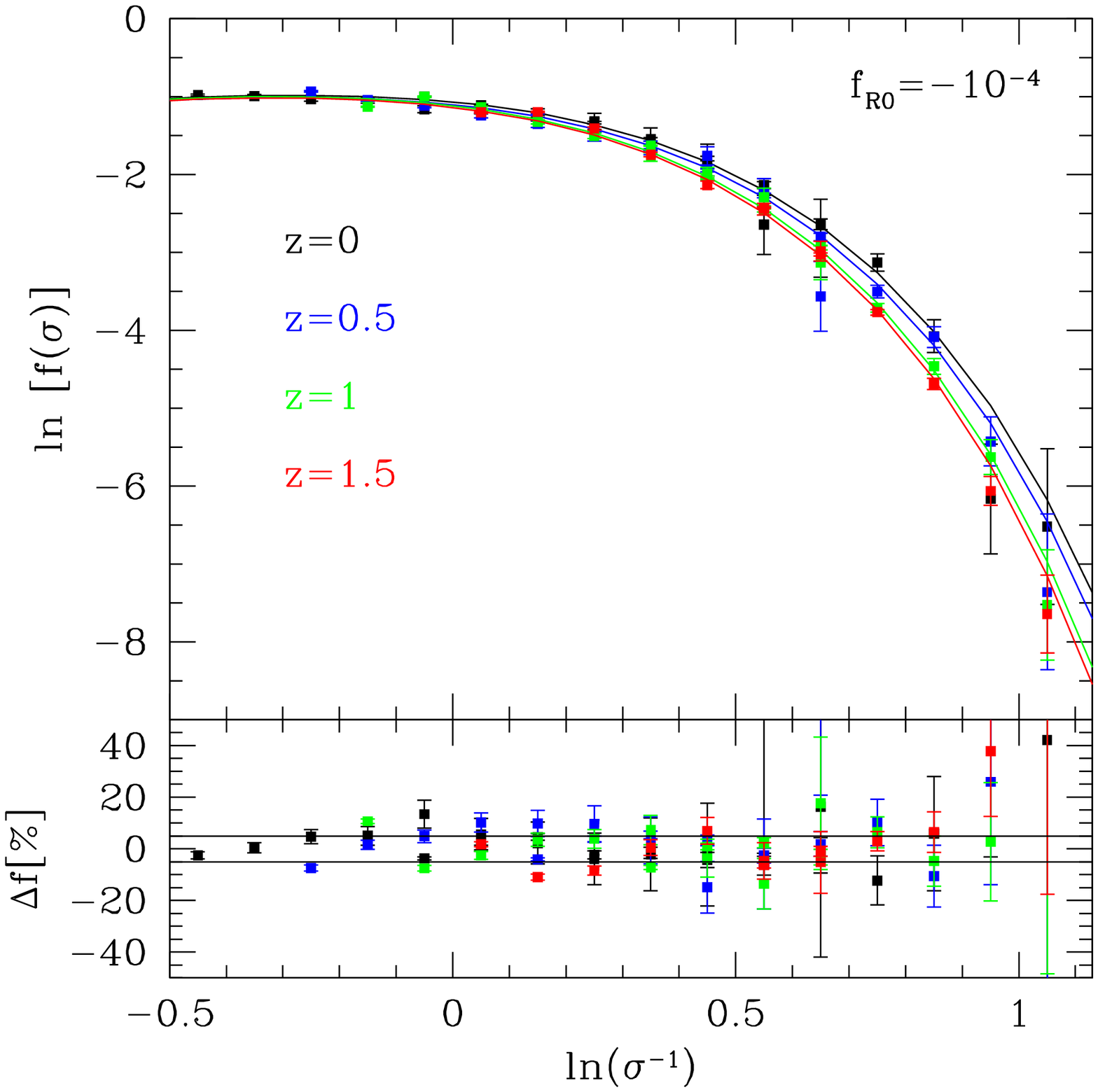}
\end{tabular}
\caption{Multiplicity function for $f_{\rm R0}=-10^{-5}$ and $f_{\rm R0}=-10^{-4}$ at various redshifts. Solid lines are theory, squares are from simulations. Lower panels show the relative difference. Black solid lines show $\pm5\%$ differences.}\label{Fig1-bis}
\end{figure}

In recent years the abundance of galaxy clusters as a cosmological probe has been firmly established \cite{Vikhlinin}\cite{Mantz}\cite{Planck_016}\cite{Mana}\cite{Rozo}. In \cite{Koppetal}, the authors have solved the spherical collapse threshold for a class of $f(R)$ modified gravity models. The linear solution extrapolated to several redshifts was used to predict the halo mass function. In this section we compare the theoretical mass function of \cite{Koppetal} to the N-body simulations of \cite{Puchweinetal}.\\

In the excursion set approach \cite{Bondetal} based upon \cite{PressSchechter1974}, the halo mass function can be expressed as

\begin{equation}
\frac{dn}{dM}=f(\sigma)\frac{\rho_m}{M^2}\frac{d\log{\sigma^{-1}}}{d\log{M}}, \label{dndm}
\end{equation}

\noindent where $f(\sigma)=2S\mathcal{F}(S)$ is the multiplicity function, $\mathcal{F}(S)$ is the so called first-crossing distribution predicted by the excursion set theory and $\rho_m$ is the mean matter density field. The variance $\sigma$ and the variance squared S are defined as

\begin{equation}
S\equiv \sigma^2(z,R(M))=\frac{1}{2\pi^2}\int \rm dk \; k^2\; P(k,z) \vert\tilde{W}\vert^2(k,R(M))
\end{equation}

\noindent where $\tilde{W}(k,R)$ is the Fourier transform of a filter function and $P(k,z)$ the redshift dependent linear matter power spectrum.  \\

The derivative of the variance with respect to the mass is sensitive to the background evolution and thus to the cosmological parameters. The multiplicity function encapsulates all the non-linear collapse of halos and is therefore the fundamental ingredient that we want to predict. The basic idea of \cite{PressSchechter1974,Bondetal} is to predict the formation of a halo when the linear overdensity $\delta$ smoothed on a scale R(M) is above a critical threshold $B$. To avoid assigning more than one mass to the halo, the scale R is chosen to be the largest one for which $\delta$ crosses the threshold $B$. The halo mass is therefore defined by the volume enclosed by  the filter function at the scale R which satisfies $\delta(R)=B$, times the mean matter density field. Assuming a random position (e.g. $\vec{x}=0$), the smoothed overdensity $\delta(R)$ is given by

\begin{equation}
\delta(R)=\frac{1}{(2\pi)^3}\int \rm d^3k \tilde{W}(k,R) \tilde{\delta}(k).
\end{equation}

For Gaussian initial conditions a top-hat filter in Fourier space and  and GR spherical collapse, one can compute exactly the probability density function of $\delta(R)<\delta_c$: $\Pi(\delta,S)$ \cite{Bondetal}. Thus, the fraction of volume collapsed into objects on scale R or larger becomes :

\begin{equation}
F(R)=1-\int_{-\infty}^{B} \Pi(\delta,S(R)),\label{coll}
\end{equation}

\noindent from which the first-crossing rate is given by $\mathcal{F}(S)=dF(R)/dS(R)$. In \cite{MR1} the authors used a path integral technique to compute $\Pi(\delta,S)$ assuming a top-hat filter in real space. This filter is convenient since it defines spherical volumes unlike a top-hat filter in Fourier space (sk-filter). In \cite{MR2}, the solution of the first-crossing rate for overdensity smoothed with a top-hat filter in real space (sx-filter) has been extended to the case of a stochastic threshold B. In this case, the barrier is no longer deterministic and is defined by a Gaussian distribution with a mean value $\bar{B}=\delta_c$ and a variance $<B(S_1)B(S_2)>=D_B\rm min(S_1,S_2)$. The same model was extended \cite{CA1,CA2,AC1} to the case where $\bar{B}=\delta_c+\beta S$. For a positive $\beta$, this barrier increases with the variance. This implies that low mass halos require a larger linear overdensity to collapse in agreement with \cite{SMT}. The analytical prediction for this model of barrier and filter (ie: top-hat in real space) was tested against the exact numerical solution \cite{CA2,ARSC} by performing Monte Carlo random walks \cite{Bondetal}. It was shown that the analytical solution reproduces the exact solution within $5\%$. Furthermore,  this barrier model predicts a halo mass function in excellent agreement with N-body simulations \cite{CA1} and is consistent with the initial conditions \cite{ARSC,AWWR} (ie: the parameters of the barrier can be measured in the initial conditions of the N-body simulation). The choice of the linear drifting term $\beta S$ over more generic functions (e.g. $\beta S^\gamma$) was motivated by the fact that the analytical solution for the multiplicity function is exact for the sk-filter. While this model is sufficient for predicting very accurate mass functions in the $\Lambda$CDM case, it needs to be extended if one considers barriers with a mean value which does not scale linearly. \\

By solving the full modified Einstein and fluid equations, \cite{Koppetal} were able to construct an approximate functional form for the spherical collapse threshold $\delta_c$ in the case of an \textit{f}(R) model. In this model, the threshold is deterministic but unlike GR, the spherical threshold now depends on the mass of the halo. The numerical fit of \cite{Koppetal} is given by:

\begin{align} \label{deltacfit}
\delta_c(z,M,f_{\rm R0})&= \nonumber
\delta^{\Lambda}_c(z) \Bigg\{ 1+ b_2 (1+z)^{-a_3} \times \\
&\left( m_{b} -\sqrt{m_{b}^2+1}\right)+ b_3(\tanh m_{b}-1) \Bigg\}\, ,\\
m_{b}(z,M,f_{\rm R0})&=(1+z)^{a_3}\times \\
&\left(\log_{10} [M/(M_\odot h^{-1})]-m_1(1+z)^{-a_4}\right) \nonumber\, ,\\
m_1(f_{\rm R0})&=  1.99 \log_{10}\left|f_{\rm R0}\right|+26.21 \nonumber\, ,\\
b_2 &= 0.0166\nonumber\, ,\\
b_3 (f_{\rm R0})&=0.0027 \cdot (2.41-\log_{10}\left|f_{\rm R0}\right| )\nonumber\, ,\\
a_3(f_{\rm R0})&=1 + 0.99 \exp\left[-2.08 (\log_{10}\left|f_{\rm R0}\right| + 5.57)^2\right]\nonumber\, ,\\
a_4(f_{\rm R0})&=\left(\tanh\left[0.69\cdot (\log_{10}\left|f_{\rm R0}\right| + 6.65) \right] + 1\right) 0.11\nonumber\, ,
\end{align}

\noindent where z is the redshift, M the mass and $f_{\rm R0}=\frac{df}{dR}\left(R_0\right)$ with $R_0$ modified gravity curvature parameter today \cite{Hu_Sawicki_2007}. Using Eq.(\ref{deltacfit}) for the spherical collapse, and keeping the drifting term $\beta$, the multiplicity function for the sx-filter (id: top-hat in real space), is given by \cite{Koppetal}:

\begin{equation}
f(\sigma)=f^{\rm GR}_{\rm sx}(\sigma) \frac{f_{\rm sk}(\sigma)}{f^{\rm GR}_{\rm sk}(\sigma)}+\mathcal{O} (\kappa^2)\label{fall}
\end{equation}

where $\kappa$ is called the non-Markovian amplitude parameter, $\kappa\sim 0.46$, and 

\begin{equation}
f^{\rm GR}_{\rm sx}(\sigma)\simeq f^{\rm GR}_{\rm sk}(\sigma)+f_{1,\beta=0}^{m-m}(\sigma)+
f_{1,\beta^{(1)}}^{m-m}(\sigma)+f_{1,\beta^{(2)}}^{m-m}(\sigma)\,,\label{ftot}
\end{equation}

with 
\begin{equation}
 f^{\rm GR}_{\rm sk}(\sigma)=\frac{\delta_c}{\sigma}\sqrt{\frac{2a}{\pi}}\,e^{-\frac{a}{2\sigma^2}(\delta_c+\beta\sigma^2)^2},\label{fsigma0}
\end{equation}

\begin{equation}
f_{1,\beta=0}^{m-m}(\sigma)=-\tilde{\kappa}\dfrac{\delta_c}{\sigma}\sqrt{\frac{2a}{\pi}}\left[e^{-\frac{a \delta_c^2}{2\sigma^2}}-\frac{1}{2} \Gamma\left(0,\frac{a\delta_c^2}{2\sigma^2}\right)\right]\,,\label{beta0}
\end{equation}
\begin{equation}
f_{1,\beta^{(1)}}^{m-m}(\sigma)=- a\,\delta_c\,\beta\left[\tilde{\kappa}\,\text{Erfc}\left( \delta_c\sqrt{\frac{a}{2\sigma^2}}\right)+ f_{1,\beta=0}^{m-m}(\sigma)\right]\,,\label{beta1}
\end{equation}
\begin{equation}
f_{1,\beta^{(2)}}^{m-m}(\sigma)=-a\,\beta\left[\frac{\beta}{2} \sigma^2 f_{1,\beta=0}^{m-m}(\sigma)+\delta_c \,f_{1,\beta^{(1)}}^{m-m}(\sigma)\right]\,.\label{beta2}
\end{equation}

\noindent with $a=1/(1+D_B)$.

Finally, $f_{sk}(\sigma)$ is the prediction of the multiplicity function for a barrier with generic mean values and a Gaussian distribution. For Eq.(\ref{deltacfit}) we have \cite{Koppetal}

\begin{equation}
f_{sk}(\sigma)\simeq\sqrt{\frac{2a}{\pi}} e^{-a\bar{B}^2/(2\sigma^2)}\frac{1}{\sigma}\Bigg(\bar{B}-\sigma^2\frac{d\bar{B}}{d\sigma^2}\Bigg)\label{eq:fskgeneric}
\end{equation}

\noindent with $\bar{B}=\delta_c(M,f_{\rm R0}) +\beta S$. Since the spherical collapse for GR is very different from the one in a $f(R)$ model, we should assume that $\beta$ and $D_B$ are not necessarily the same for GR and $f(R)$ models. In both cases there is no theory available to predict these parameters. In \cite{ARSC,AWWR} these parameters were measured in the initial conditions. In this work we adopt the approach of \cite{CA1,CA2} to determine $\beta$ and $D_B$. We compute the best fitting values of $\beta$ and $D_B$ which reproduce the multiplicity function of the N-body simulations. We start by comparing $f_{sx}^{GR}(\sigma)$ to the N-body simulation of \cite{Puchweinetal} where halos are identified using a Friend-of-Friend (FoF) algorithm with linking length $b=0.2$. In order to simplify the analysis, we assume that for the $\Lambda$CDM case, the multiplicity function is universal. Thus for all redshifts we find that the best fitting values are $\beta=0.05$ and $D_B=0.34$. Then we recompute a best fit for $\beta$ and $D_B$ in Eq.(\ref{eq:fskgeneric}), which are specific to the modified gravity model. These values are reported in Table \ref{Tab1}.

\begin{tiny}
\begin{table}
\centering
\begin{tabular}{|l|c|c|c|}
\hline  & $f_{R0}=-10^{-4}$ & $f_{R0}=-10^{-5}  $& $ f_{R0}=-10^{-6} $\\ 
\hline & $\beta,D_B$ & $\beta,D_B$ & $\beta,D_B$ \\
\hline $z=0.0 $ & $0.05\;, 0.3$ & $0.06\;, 0.3$  &$0.08\;, 0.42$ \\ 
\hline $z=0.5$ & $0.08 \;, 0.29$ & $0.06\;, 0.26$ & $ 0.08\;, 0.42$ \\ 
\hline  $z=1.0$ & $0.08\;, 0.22$ &  $0.06 \;, 0.25$ &  $ 0.09 \;, 0.4$ \\ 
\hline  $z=1.5$ &$0.1 \;, 0.2$ &  $0.09 \;, 0.23$ &  $ 0.09\;, 0.4$ \\ 
\hline 
\end{tabular}
\caption{values of $\beta,D_B$ in Eq.(\ref{eq:fskgeneric}) for different redshifts and $f_{R0}$ parameters. For the $GR$ multiplicity function Eqns.(\ref{ftot},\ref{fsigma0}) we neglect the non-universal behaviour of the multiplicity function: for all redshifts we set $\beta=0.05$  and $D_B=0.34$ } 
\label{Tab1}
\end{table}
\end{tiny}

Note that on average $\beta$ tends to increases with redshift while $D_B$ tends to decrease with redshift in agreement with \cite{AWWR}. This means that the threshold becomes more deterministic at high redshift. Using Eq.(\ref{fall}) with the values of Tab \ref{Tab1} we obtain the prediction shown in Figs. \ref{Fig1}-\ref{Fig1-bis}. The multiplicity functions measured in the N-body simulations are represented by the squares and Eq.(\ref{fall}) by the solid lines. Lower panels show the relative difference between theory and N-body simulations. The black lines mark the $\pm5\%$ deviation. As we can see, for the various $f_{R0}$ parameters and at different redshifts, our prediction remains very accurate over all ranges of mass. 

\begin{figure}[ht]
\centering
\begin{tabular}{cc}
\includegraphics[scale=0.35]{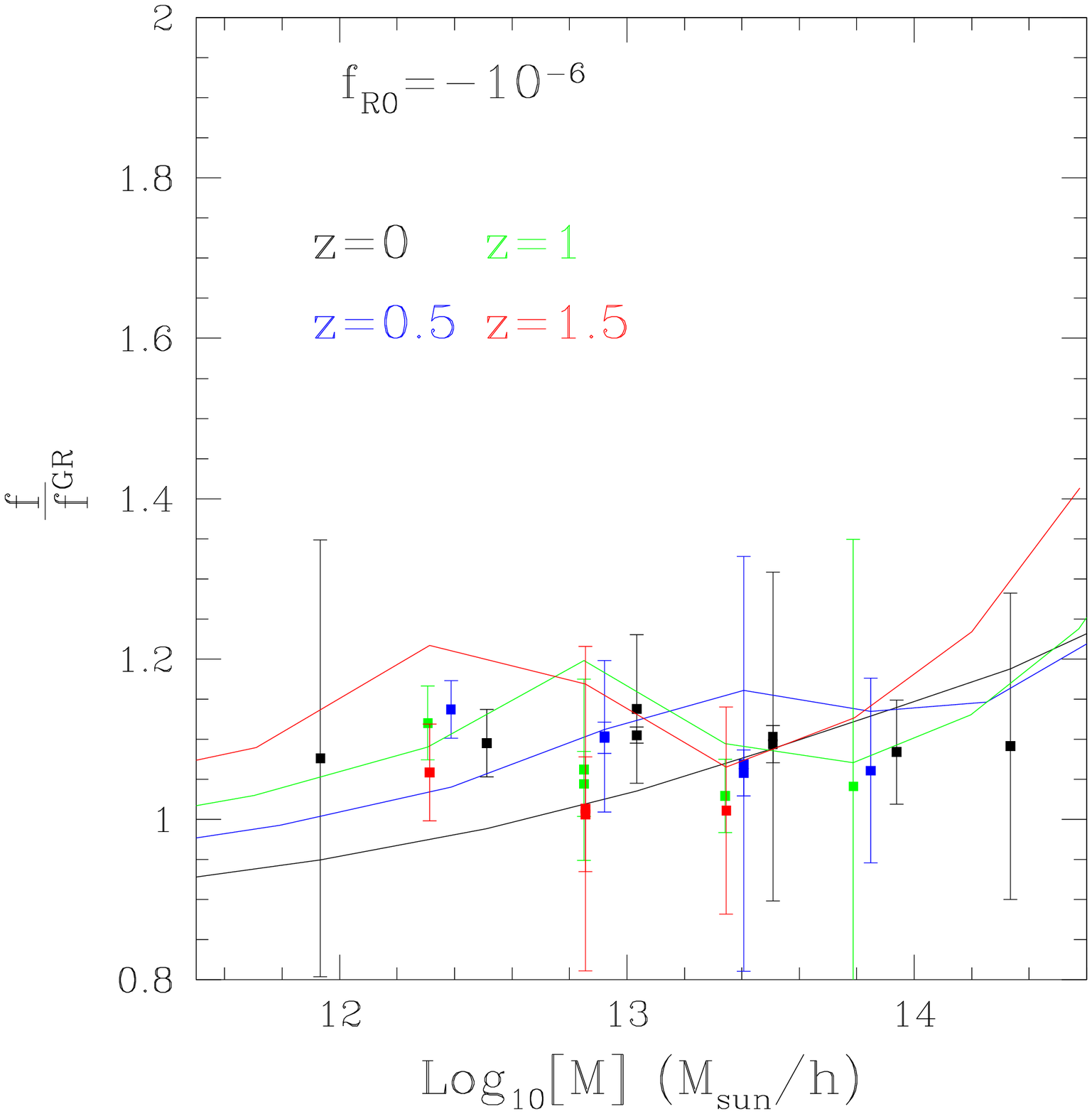}\\
\includegraphics[scale=0.35]{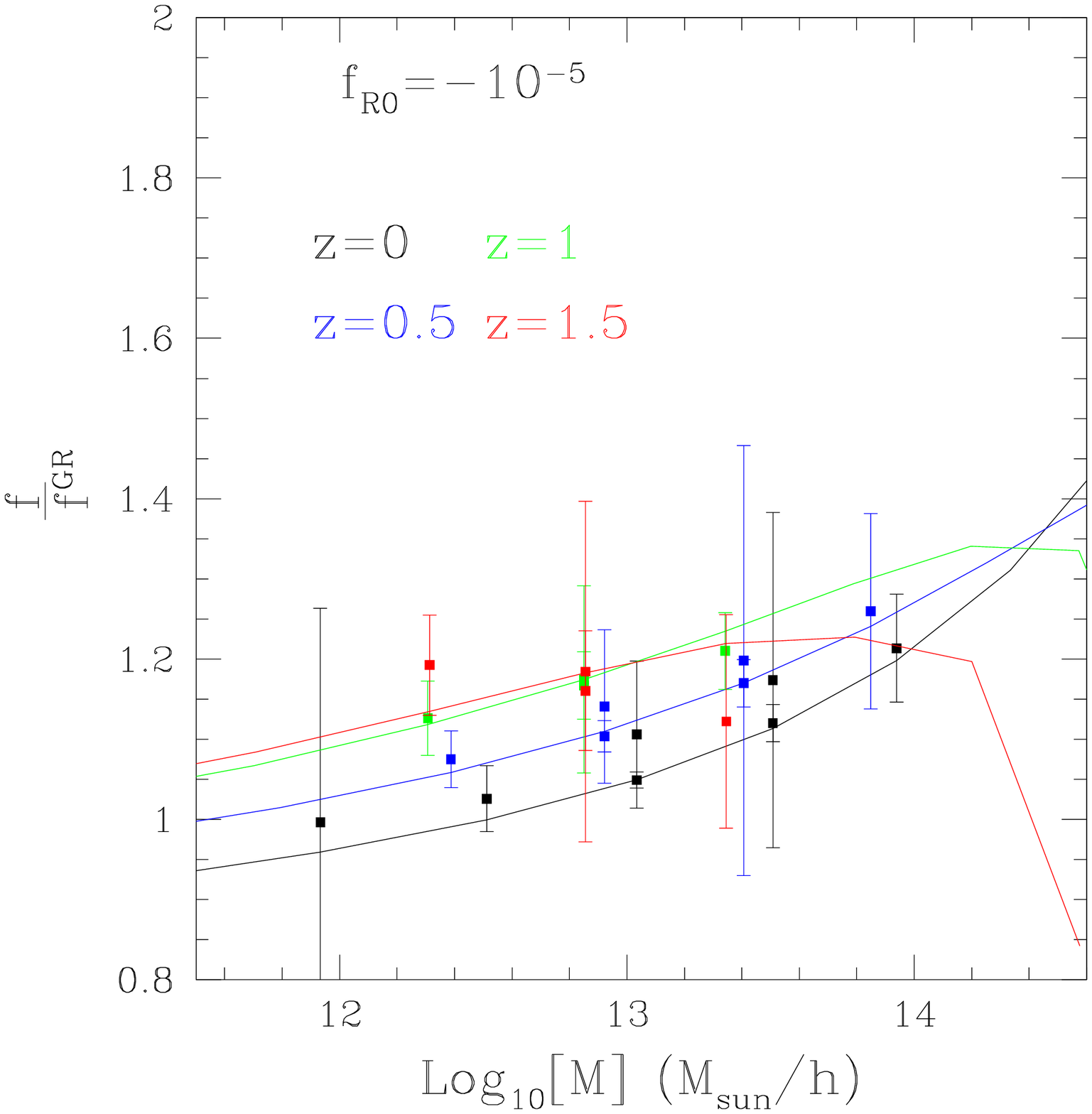}
\end{tabular}
\caption{Modified gravity multiplicity function for $f_{R0}=-10^{-6}$ and $f_{R0}=-10^{-5}$ with respect to $\Lambda$CDM multiplicity function. The solid lines show the theory while  the squares show the N-body simulation result. }\label{Fig2}
\end{figure}

\begin{figure}[ht]
\centering
\begin{tabular}{cc}
\includegraphics[scale=0.35]{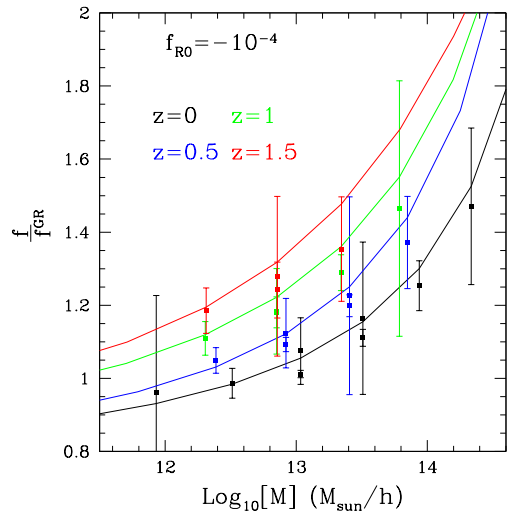}
\end{tabular}
\caption{Modified gravity multiplicity function with respect to $\Lambda$CDM multiplicity function for $f_{R0}=-10^{-4}$. The solid lines show the theory while  the squares show the N-body simulation result. }\label{Fig2bis}
\end{figure}

The deviation of the halo mass function between $\Lambda$CDM and the $\textit{f}(R)$ model is discussed in \cite{Puchweinetal}. In Figs. \ref{Fig2}-\ref{Fig2bis} we show the ratio between the $\textit{f}(R)$  and the $\Lambda CDM$ multiplicity functions. The solid lines are the theoretical ratio using the multiplicity functions of Fig. \ref{Fig1} while the squares are the ratio from the N-body simulations. Different colours correspond to different redshifts according to the legend. Typically the difference between GR and f(R)increases with mass. For $f_{R0}=-10^{-4}$ and over the mass ranges probed by the N-body simulations, the difference reaches $\sim 50 \%$. For $f_{R0}=-10^{-5}, -10^{-6}$ the difference is $\sim 30\%, 20\%$ respectively. The bumps present for the $f_{R0}=-10^{-6}$ are specific to the trend of the spherical collapse model given by Eq.(\ref{deltacfit}). Similar trends are shown in \cite{LLKZ}.

\section{Conditional mass function and Linear halo bias for modified gravity}\label{secCondi}
The linear halo bias measures how halos are biased compared to the underlying dark matter density field on large scales. It can be measured by taking the square root of the ratio between the 2-point correlation function of halos and the one of the underlying matter density field. The peak-background split approach \cite{Bardeenetal,Coleetal,Moetal,ST99} provides a theoretical framework to compute it. The idea is to measure the abundance of halos formed at an initial overdensity $\delta_0(r_0)$ relative to the unconditional mass function. Assuming $r_0\rightarrow \infty$ (e.g. $\delta_0\ll 1$) then the linear halo bias is given by

\begin{figure}[ht]
\centering
\begin{tabular}{cc}
\includegraphics[scale=0.35]{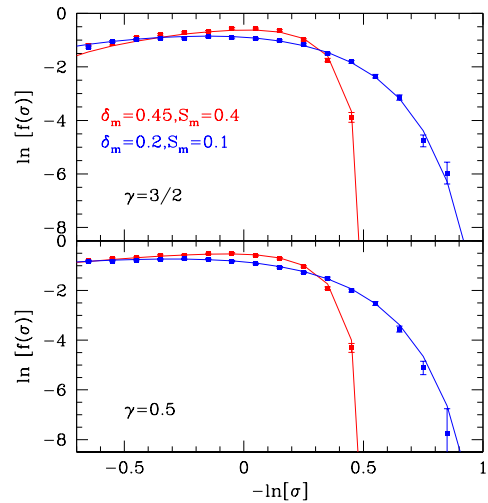}
\end{tabular}
\caption{$f(\sigma)$ from Eqn.(\ref{genericcondi}) is plotted as a solid line and  the exact Monte Carlo solution is shown by squares for different conditions $(\delta_m,S_m)$ (red and blue). The upper panel correspond to a barrier with parameters $(\delta_c=1.673,\beta=0.1,\gamma=0.5)$ while the lower panel is for $(\delta_c=1.673, \beta=0.1,\gamma=1.5)$. }\label{Fig3}
\end{figure}

\begin{equation}
b_h=1+\frac{1}{\mathcal{F}(S|\delta_0=0,r_0=0)}\frac{\partial \mathcal{F}(S|\delta_0,r_0=0)}{\partial \delta_0}\mid_{\delta_0=0}\label{bhtheo}
\end{equation}

In this section we  compute the conditional mass function for a diffusing barrier with generic drift. We first consider the general case without taking any limit for the perturbations $\delta_0$. Then we derive the linear bias and test the robustness of our prediction using Monte Carlo random walks\cite{Bondetal}. \\

In \cite{MR1}, the authors have developed a path-integral framework where the excursion set approach can be reformulated for generic filters. By discretizing the variable $S$ in step $\Delta S$ such that $S_k=k\Delta S$ with $k=1,..,n$, the probability density (PDF) of finding an overdensity $\delta$ on a scale $S(R)$, without crossing the threshold B on a scale $S'<S$ is given by 

\begin{equation}
\Pi(\delta_0,\delta_n,S_n)=\int_{-\infty}^{B} d\delta_1 ... \int_{-\infty}^{B} d\delta_{n-1} W(\delta_0,..\delta_n,S_n)
\end{equation}
where

\begin{equation}
W(\delta_0,...\delta_n,S_n)=\int  \mathcal{D}\lambda e^{i\sum_{i=1}^n \lambda_i \delta_i}\langle e^{-i \sum_{i=1}^n \lambda_i \delta(S_i)}\rangle.
\end{equation}

The measure $\mathcal{D}\lambda\equiv d\lambda_1/(2\pi)...d\lambda_n/(2\pi)$ and the brackets $\langle ...\rangle$ refer to an ensemble average of the random walks. This is the well-known \textit{exp} function of the partition function used in particle physics. The fraction of collapsed halos is also derived using Eq.(\ref{coll}) once the PDF is known. In \cite{Maetal} the probability to get $\delta_n$ on scale $S_n$ knowing that the smoothed field had a value $\delta_m$ on scale $S_m$ is

\begin{equation}
\begin{split}
&P(\delta_n,S_n\vert \delta_m,S_m)=\\
&\frac{\int_{-\infty}^{B} d\delta_1 ...\hat{d\delta_m}... \int_{-\infty}^{B} d\delta_{n-1} W(\delta_0,...,\delta_m,...\delta_n,S_n)}{\int_{-\infty}^{B} d\delta_1 ... \int_{-\infty}^{B} d\delta_{m-1} W(\delta_0...\delta_m,S_m)}\label{PDFcondi}
\end{split}
\end{equation}
where $\hat{d\delta_m}$ means we do not integrate over $\delta_m$. Therefore the first-crossing distribution is

\begin{equation}
\mathcal{F}(S|\delta_m,S_m)=-\int_{-\infty}^{B} \frac{\partial P(\delta_n,S_n|\delta_m,S_m)} {\partial S_n}.
\end{equation}

For a sharp-k filter (sk) Eq.(\ref{PDFcondi}) is greatly simplified \cite{Maetal,CA2}. The Markovian property implies that 
\begin{equation}
\begin{split}
&P(\delta_n,S_n|\delta_m,S_m)=\\
&\int_{-\infty}^{B} d\delta_{m+1}...\int_{-\infty}^{B} W(\delta_m ...\delta_n,S_n-S_m) 
\end{split}
\end{equation}\label{condisk}

The random walk analogy for this equation implies the system does not memorize previous steps, such that the probability to pass by $\delta_m(S_m)$ is the same as starting the random walk on scale $S_m$ with initial values $\delta_m$ instead of $(\delta_0=0,S_0=0)$. This leads to the rescaling $\delta_0\rightarrow \delta_m$, $S\rightarrow S-S_m$ in the multiplicity function. This is exact for constant threshold B, as in the GR spherical collapse case. In the case of a diffusing barrier with a sk filter, we should carefully rewrite the previous equations. In the model of \cite{CA2,ARSC} the barrier is not correlated with the smoothed field ($<B(S)\delta(S')>=0$). The barrier PDF is defined by 
\begin{equation}
\Pi_B(B_0,B_n,S_n)=\frac{1}{\sqrt{2\pi D_B S}}e^{-\frac{(B-\bar{B})^2}{2 D_B S}}.\label{pdfb}
\end{equation}

We assume first $\bar{B}=B_0$, a constant, such that the barrier PDF satisfies the Fokker-Planck Equation $\partial \Pi_B/\partial S= \frac{D_B}{2} \partial^2 \Pi_B /\partial B^{2}_n$. The first-crossing distribution then becomes

\begin{equation}
\begin{split}
&\mathcal{F}(S|\delta_m,S_m)=\\
&-\frac{\partial }{\partial S}\int_{-\infty}^{\infty}  dB\; \Pi_B(B_0,B_n,S) \int_{-\infty}^{B}d\delta \;\Pi(\delta_m,\delta_n,S_m-S_n). \label{theofc}
\end{split}\, .
\end{equation}

The PDF $\Pi(\delta_m,\delta_n,S_m-S_n)$ is a solution of the Fokker-Planck Equation: $\partial \Pi/\partial S= \frac{1}{2} \partial^2 \Pi /\partial \delta^{2}_n$. Furthermore the solution must satisfy the absorbing boundary condition $\Pi(\delta_m,\delta_n=B,S_m-S_n)=0$ and the initial condition that $\Pi(\delta_m,\delta_n,S_m-S_n=0)=\delta_D(\delta_n-\delta_m)$, leading to the solution \cite{Bondetal} 

\begin{equation}
\begin{split}
&\Pi(\delta_m,\delta_n,S_n-S_m)=\\
&\frac{1}{\sqrt{2 \pi (S_n-S_m)}} \lbrace e^{-\frac{(\delta_n-\delta_m)^2}{2(S_n-S_m)}} - e^{-\frac{(2 B-\delta_n-\delta_m)^2}{2(S_n-S_m)}}\rbrace. \label{pdfdel2}
\end{split}
\end{equation}

Plugging Eq.(\ref{pdfdel2}) into Eq.(\ref{theofc}) leads to 
\begin{equation}
\begin{split}
&\mathcal{F}(S|\delta_m,S_m)=\\
&-\frac{\partial }{\partial S}\int_{-\infty}^{\infty}  dB\; \Pi_B(B_0,B_n,S)\; \rm Erf\left[ \frac{B-\delta_m}{\sqrt{2(S-S_m)}}\right] 
\end{split}
\end{equation}

The derivative with respect to S leads to two terms: the derivative of the error function term and the derivative of the term in Eq.(\ref{pdfb}).  Using of the Fokker-Planck equation and Eq.(\ref{pdfb}) we perform an integration by parts leading to

\begin{equation}
\mathcal{F}(S|\delta_m,S_m)=\sqrt{\frac{1}{2\pi}}\frac{(B_0-\delta_m)(1+D_B)}{[(1+D_B)S-S_m]^{3/2}}e^{-\frac{(B_0-\delta_m)^2}{2(1+D_B)S-2S_m}}\, .\label{solint}
\end{equation}

From this expression we can see that setting $D_B=0$ leads to the original solution of \cite{Bondetal}. This solution for $\delta_m=0$ is also in agreement with \cite{Maetal} who use for the barrier in Eqn.(\ref{pdfb}) with $\bar{B}=\delta_c$. For modified gravity we want to extend Eq.(\ref{solint}) to any generic mean value $\bar{B}$ so that we can use Eq.(\ref{deltacfit}), which is no longer a constant but a function of the mass $M(S)$. For a generic barrier, the first-crossing does not have an exact analytical solution for the unconditional and conditional mass function. This is due to the absorbing boundary condition, which can not be fulfilled except for a constant or a linear barrier. In \cite{Koppetal}, the authors computed an approximation for the unconditional multiplicity function with a generic barrier . In \cite{Koppetal}, for the functional trend of Eq.(\ref{deltacfit}), the authors found that Eq.(\ref{eq:fskgeneric}) reproduces the exact Monte-Carlo solution within $5\%$. In what follows we propose to extend this approximation to the conditional mass function by taking 

\begin{equation}
\begin{split}
&f(S|\delta_m,S_m)\simeq\sqrt{\frac{2}{\pi}} \left[ \bar{B}(S)-S\frac{d\bar{B}}{dS}+\frac{S^2}{2}\frac{d^2 \bar{B}}{dS^2}-\delta_m)\right] \times\\
&\frac{S (1+D_B)}{[(1+D_B)S-S_m]^{3/2}}e^{-\frac{(\bar{B}(S)-\delta_m)^2}{2(1+D_B)S-2S_m}}\label{genericcondi}\,.
\end{split}
\end{equation}

In order to test the robustness of our expression we ran a series of Monte Carlo random walks\cite{Bondetal} and solved the exact multiplicity function for various conditions ($\delta_m,S_m$). In particular we show in Fig. \ref{Fig3} the case where  $D_B=0.4$ and $\bar{B}=1.673+\beta S^{\gamma}$ with $\gamma=0.5$ in the upper panel and $\gamma=3/2$ in the lower panel, both with $\beta=0.1$. 
For every test we performed (including the case where $\delta_c$ is given by Eq.(\ref{deltacfit}) ), a good agreement between the exact solution and Eq.(\ref{genericcondi}) has been achieved.

In the case where $\bar{B}=\delta_c+\beta S^\gamma$, with $\gamma\leq 1$, the second derivative in Eq.(\ref{eq:fskgeneric}) can be omitted. This is the case for the modified gravity model we consider here, in the limit of small perturbation $\delta_m\rightarrow 0$. \\

From Eqs.(\ref{genericcondi},\ref{bhtheo})  we obtain the following linear halo bias in Eulerian space:

\begin{equation}
b_h(S)=1+\left\lbrace \frac{\bar{B}}{S(1+D_B)} -\frac{1}{\bar{B}-S\frac{d\bar{B}}{dS}}\right\rbrace \label{linbias}\, ,
\end{equation}

This expression is valid for a drifting diffusive barrier defined by Eq.(\ref{pdfb}) and in particular for $\bar{B}=\delta_c(S)+\beta S$, with $\delta_c$ defined by Eq.(\ref{deltacfit}). The parameters $\beta$ and $D_B$ are fixed by the halo mass function.

\section{ $f(R)$ halo profiles }\label{secPro}

\begin{figure*}[ht]
\centering
\includegraphics[width=\textwidth]{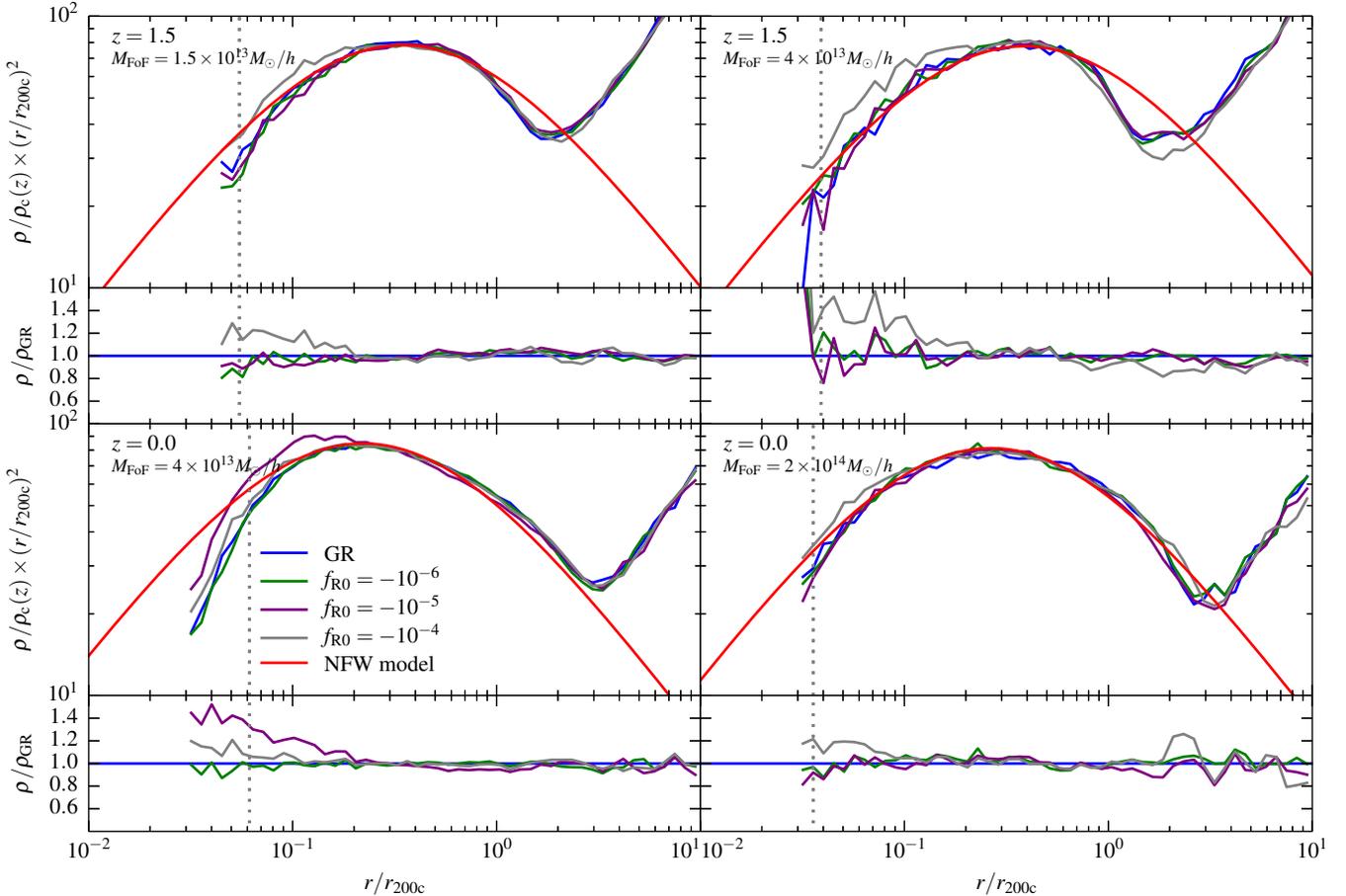}
\caption{Density profiles of halos in simulations with GR and $f(R)$. We also compare to the NFW profile obtained from the halo concentration model. The profiles in the larger panels are given in units of the critical density at the considered redshift and have been multiplied by $(r / r_{200\mathrm{c}})^2$ to reduce the dependence on radius and allow a more accurate comparison. When plotted in this way the maximum of a NFW profile is attained at the scale radius. The relative differences of the $f(R)$ simulations compared to the GR simulation are shown in the smaller panels. All radii are shown in units of $r_{200\mathrm{c}}$. Results are plotted for $z=1.5$ (upper panels) and $z=0$ (lower panels), as well as for different halo mass bins. For the simulations, the mean profiles of all halos with a FoF mass within 20 \% of the value indicated in each panel are shown. The NFW profiles have been computed for the quoted FoF mass values. Vertical dotted lines show the gravitational softening in the simulations.}\label{Fig_halo_profiles}
\end{figure*}

The structure of dark matter halos is well-known in GR (e.g. \cite{Coleetal2}\cite{Navarroetal}\cite{Pradaetal}). Their profiles are in good agreement with the NFW fit \cite{Navarroetal}, given by 

\begin{equation}
\rho(r)=\frac{\rho_s}{(r/r_s)(1+r/r_s)^2}\label{eq:NFW}\, ,
\end{equation}
where $\rho_s$ paramterizes the amplitude of the density profile and the scale radius $r_s$ characterizes where the logarithmic slope of the density profile changes.

In Fig.~\ref{Fig_halo_profiles}, we compare the halo density profiles in our $f(R)$ simulations to the profiles obtained for GR. Results are shown for halos of different mass and at different redshifts. The radii are given in units of $r_{200\rm crit}$, the spherical overdensity radius within which the average density is 200 times the critical density of the Universe. The mass enclosed in this radius is called $M_{200\rm crit}$. Overall the differences at fixed $M_{200\rm crit}$ are rather minor, except maybe for the very central region. For $r>r_{200\rm crit}$ the deviations in the density profile between $f(R)$ and GR are typically $<10\%$ for all considered values of $f_{R0}$. Near the centre, for $r \lesssim r_{200\rm crit}$ deviations can be larger and reach in some cases a density increase of up to $\sim 40\%$ compared to GR. The $f_{R0}$ value for which the largest deviations are observed seem to depend on halo mass and redshift. This is consistent with the findings of \cite{Hammamietal2015}. The observed trend suggests that the largest deviations in the profile shape from GR are observed for objects near the screened/unscreened transition as has also been found for the velocity dispersion profiles \cite{Gronkeetal2015}.

In order to use the profiles in a halo model for the matter power spectrum, we need to be able to convert the halo mass to the profile parameters $\rho_s$ and $r_s$. As the effects of $f(R)$ on the profiles at fixed mass are largely confined to the very central region which only contribute very little to the matter power spectrum at the wavenumbers we consider here \cite{vanDaalenSchaye2015}, we choose to use the same mass-dependent NFW fits for both GR and $f(R)$. The $f(R)$ effect on the matter power spectrum is then given by the difference in the halo mass function.

A very useful quantity for computing the profile parameters is the halo concentration, which can be defined by $c_{200\rm crit} \equiv r_{200\rm crit} / r_s$. In the following we use the model for halo concentration as a function of halo mass from \cite{Maccioetal2008}, which is a modified version of the model presented in \cite{Bullocketal2001}. Predictions of this model are shown to compare well to LCDM simulations of cosmic structure formation in different cosmologies and at different redshifts \cite{DuttonMaccio2014}. The main steps in this model for obtaining the concentration from $M_{200\rm crit}$ are

\begin{itemize}

\item Defining a characteristic mass $M_*$, whose assembly redshift relates to the concentration. Here a constant fraction of the viral mass, $M_* = F \times M_{200\rm crit}$, is assumed. In the following we use $F=0.01$ in agreement with \cite{Maccioetal2008}.

\item Based on the spherical collapse model a collapse, redshift $z_c$ is found which satisfies $\sigma(z_c, R(M_*)) = \delta_c$. As the concentration is not particularly sensitive to the exact value of the collapse threshold, we here assume $\delta_c=1.686$ for simplicity. 

\item The concentration is then obtained by $c_{200\rm crit}(z) = K_{200\rm crit} \times (\frac{\rho_{\rm crit}(z_c)}{\rho_{\rm crit}(z)})^{1/3}$.

\end{itemize}

As illustrated in Fig.~\ref{Fig_halo_profiles}, we obtain good agreement with the density profiles in our simulations for $K_{200\rm crit} = 3.4$. The same value was found to fit a simulation based on a WMAP3 cosmology well in \cite{Maccioetal2008}.

Once $c_{200\rm crit}$ is known for a given $M_{200\rm crit}$, the corresponding $\rho_s$ and $r_s$ can then be obtained by considering
\begin{equation}
  M_{200\rm crit} = \int_{0}^{r_{200\rm crit}} dr\; 4\pi r^2 \rho(r) = 4\pi\frac{\rho_s r_s^3}{g(c_{200\rm crit})}, \label{eq:m200_c}
\end{equation}
where
\begin{equation}
   g(c)\equiv\frac{1}{\ln(1+c)-c/(1+c)}.
\end{equation}
Using $M_{200\rm crit} = 200 \rho_{\rm crit}(z) \frac{4}{3}\pi r_{200\rm crit}^{3}$, we find
\begin{equation}
  \rho_s = 200 \rho_{\rm crit}(z) \frac{g(c_{200\rm crit}) c_{200\rm crit}^3}{3},
\end{equation}
while the scale radius is simply given by
\begin{equation}
  r_s = \frac{r_{200\rm crit}}{c_{200\rm crit}}.
\end{equation}
This then completely defines the NFW density profile. 

The halo concentration model presented above is based on the spherical over density mass $M_{200\rm crit}$ rather than the FoF mass that we have used in section \ref{secMF}. To convert between these different mass definitions, we need to make use of the boundary density of FoF groups, which for a linking length of $b=0.2$ is given by $\Delta_b \approx 82$ times the mean matter density \cite{Moreetal2011}. Assuming spherical symmetry we then obtain the boundary radius, $r_{\rm FoF}$ of the FoF group from $\rho(r_{\rm FoF}) = \Delta_b \times \rho_{\rm mean}(z)$. Inserting Eqn.~(\ref{eq:NFW}) and solving for $r_{\rm FoF}$ we find
\begin{equation}
   r_{\rm FoF} = \frac{r_s}{3} (-2 + x + \frac{1}{x}),
\end{equation}
where
\begin{equation}
  x = \frac{2^{1/3}}{(2 + 27 y + 3 \sqrt{12 y + 81 y^2})^{1/3}},
\end{equation}
and
\begin{equation}
  y = \frac{\rho_s}{\Delta_b \rho_{\rm mean}(z)}.
\end{equation}
In analogy to Eq.~(\ref{eq:m200_c}), the FoF mass can then be computed by
\begin{equation}
  M_{\rm FoF} =  4 \pi \frac{\rho_s r_s^3}{g(\frac{r_{\rm FoF}}{r_s})}.
\end{equation}

In practice, we proceed in the following way to obtain the profile of a halo with a given FoF mass. We start with a grid of $M_{200\rm crit}$ values that spans the relevant range. For each value, we then compute $c_{200\rm crit}$, $\rho_s$, $r_s$ and $M_{\rm FoF}$ as detailed above. In the list that we obtain in this way, we can then interpolate in $M_{\rm FoF}$ to get the NFW parameters $\rho_s$ and $r_s$ for any given FoF mass.

\section{Halo model: non-linear matter power spectrum}\label{secHM}
\begin{figure*}[ht]
\begin{center}$
\begin{array}{cc}
\includegraphics[scale=0.45]{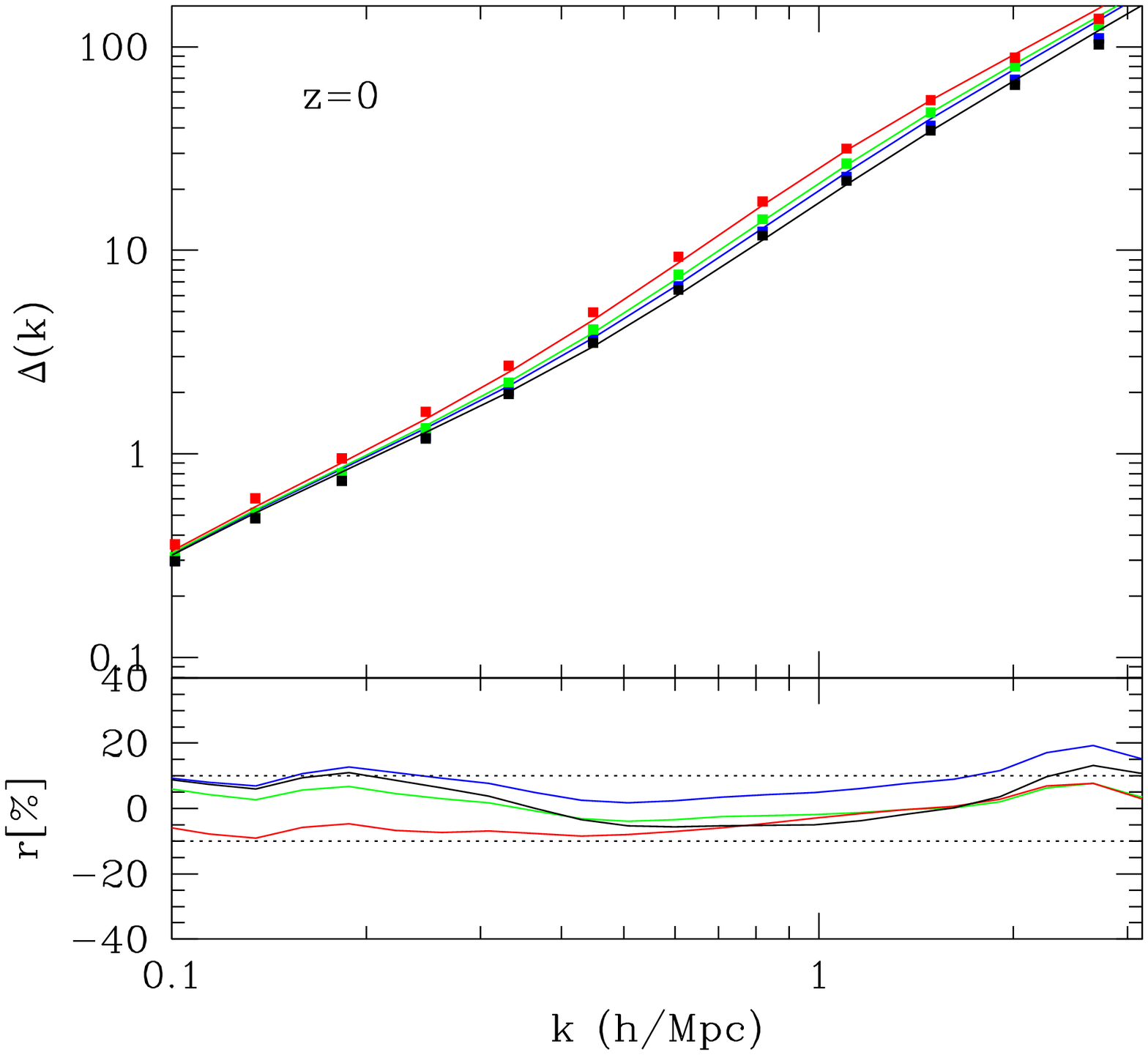} &
\includegraphics[scale=0.45]{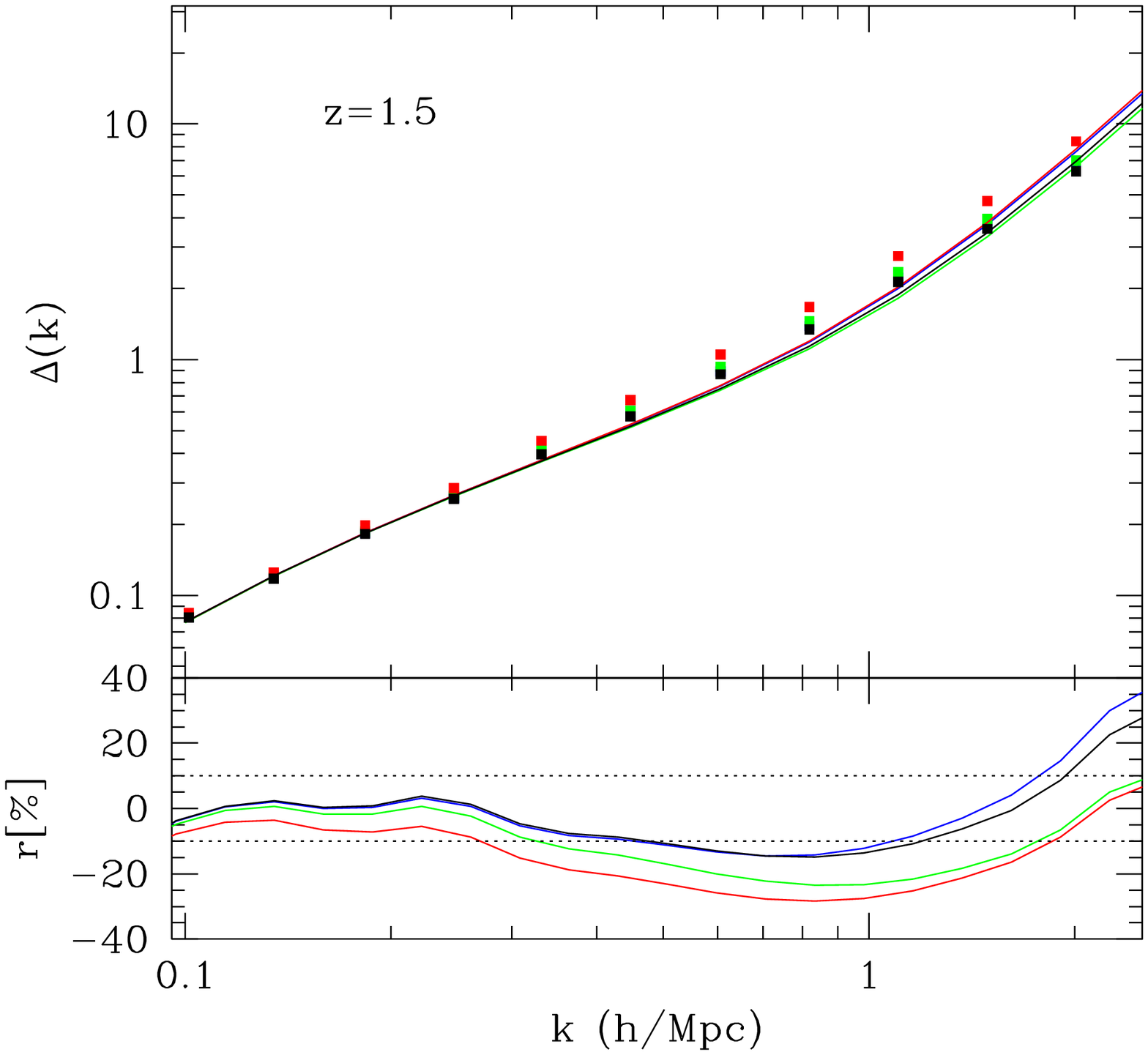}
\end{array}$
\end{center}
\caption{In the upper panel is the measured non-linear power spectrum at $z=0$ for GR (black), $f_{R0}=-10^{-6}$ (blue), $f_{R0}=-10^{-5}$ (green) and $f_{R0}=-10^{-4}$ (red) plotted in squares and the halo model prediction (solid lines). Lower panel: relative difference between the measured and predicted $\Delta(k)$ at $z=0$ (upper panel) and $z=1.5$ (lower panel).}\label{pkz0all}
\end{figure*}

In this section we use the halo model \cite{CooraySheth} to compute the non-linear power spectrum for the  modified gravity $f(R)$ model. We test our prediction with N-body simulations. % and MGcamb \cite{MGcamb}. \\
In the halo model, the non-linear power spectrum is given by 

\begin{equation}
P(k)=P^{1h}(k)+P^{2h}(k),
\end{equation}

where $P^{1h},P^{2h}$ are the one- and two-halo terms:

\begin{equation}
P^{1h}(k)=\int dm \frac{d\;n(m)}{dM} \left( \frac{m}{\bar{\rho}}\right)^2 \vert u(k\vert m)\vert^2,  
\end{equation}

\begin{equation}
P^{2h}(k)\simeq \left[ \int dm \frac{d\;n(m)}{dM} \left( \frac{m}{\bar{\rho}}\right)u(k\vert m) b(m)\right]^2  P^{lin}(k).   
\end{equation}

The $u(k\vert m)$ is the Fourier transform of the dark matter distribution within a halo of mass m, (see Eq.80 in \cite{CooraySheth}),  dn(M)/dM is the halo mass function, $b(M)$ is the linear halo bias and $P^{lin}(k)$ the linear dark matter power spectrum. These equations assume that at any redshift, all dark matter is contained in halos. This is not exactly the case for the $\Lambda$CDM model, especially for non-spherical collapse \cite{CA2}.

In order to predict the non-linear power spectrum, we use the halo mass function in Eq.(\ref{dndm}) with the multiplicity function of Eq.(\ref{ftot}) calibrated with the parameters $\delta_c(M),\beta,D_B$ given by Eq.(\ref{deltacfit}) and Table \ref{Tab1}. Those parameters also set the linear Eq.(\ref{linbias}). For the profile of halos, we use the GR values of the concentration parameters in the NFW profile that we computed in Sec.\ref{secPro}.  

In addition, due to the numerical limitation of the integrated mass range, we ensure that the halo term with the largest contribution to the power spectrum equals the linear theory expression on large scale by renormalizing the linear bias with a factor q.  Following the procedure of \cite{Scocci2001}\cite{Pielorzetal}, we compute q as follow:

\begin{equation}
 \int_{m_{min}}^{m_{max}} dm \; m\; \frac{n(m)}{\bar{\rho}} \frac{b(m)}{q}=1. 
\end{equation}

In Fig.~\ref{pkz0all} we can see the power spectrum at $z=0$ obtained from the N-body simulations for $f_{R0}=-10^{-4},-10^{-5},-10^{-6}$ and GR shown by the red, green, blue and black squares respectively. The associated solid lines show the halo model predictions. The relative differences between theory and N-body simulations are shown on the lower panel of Fig~\ref{pkz0all}. Our predictions for the halo model lead to $\sim 10\%$ accuracy with respect to N-body simulations at  $z=0$. Similarly to previous studies \citep{Lombriser2014,Brax2013, Zhao}, the accuracy of the predictions decreases for higher redshift up to $\sim 35 \%$ at $z=1.5$ for $f_{R0}=-10^{-4}$ and $\sim 20 \%$ for GR. This can be seen in the lower panel of Fig.~\ref{pkz0all}.

% check out refRecently, cite{Ishradetal} have developed an extansion of the halo model by...
Overall we note that the $f(R)$ models increase the power, particularly on small spatial scales $k \gtrsim 0.5 h {\rm Mpc}^{-1}$. For $k=1 h {\rm Mpc}^{-1}$, the relative difference to GR reaches up to $\sim 45 \%$ at $z=0$, ($\sim 30\%$ at z=1.5). Our results are similar to those found in \citep{Li2013,Zhao,Lombriser2014}. We note that our predictions could potentially be improved using a halo profile which would reproduce low mass halo profiles more accurately at high redshift, (see recent work \cite{Ishram} in this perspective). However, as we already mention, the halo model assumed all the dark matter to be contained inside halos. This approximation is worse for higher redshift, (e.g. \cite{Lazanu}).

\section{Void abundances}\label{secVA}

\begin{figure}[ht]
\centering
\includegraphics[width=\columnwidth]{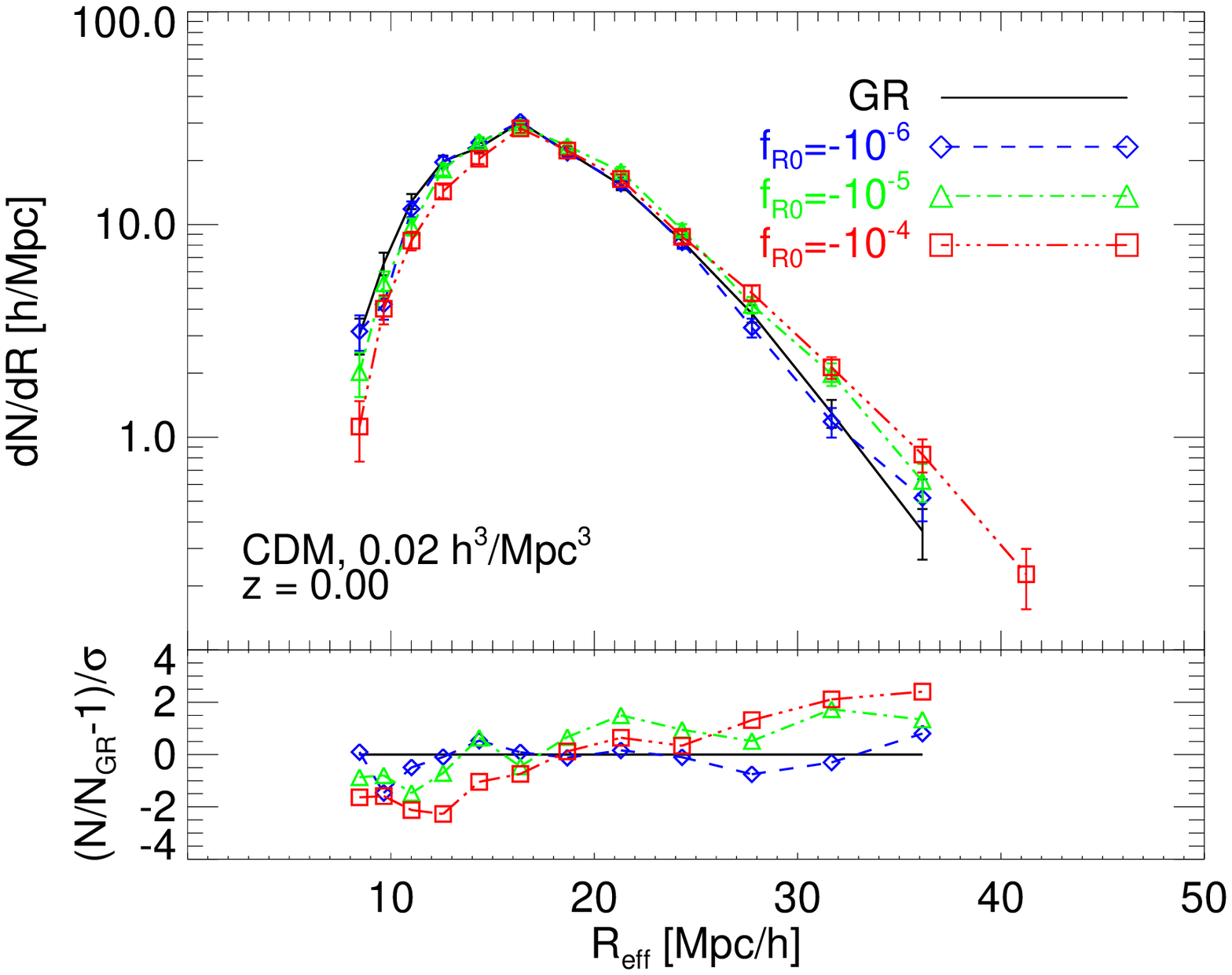}
\includegraphics[width=\columnwidth]{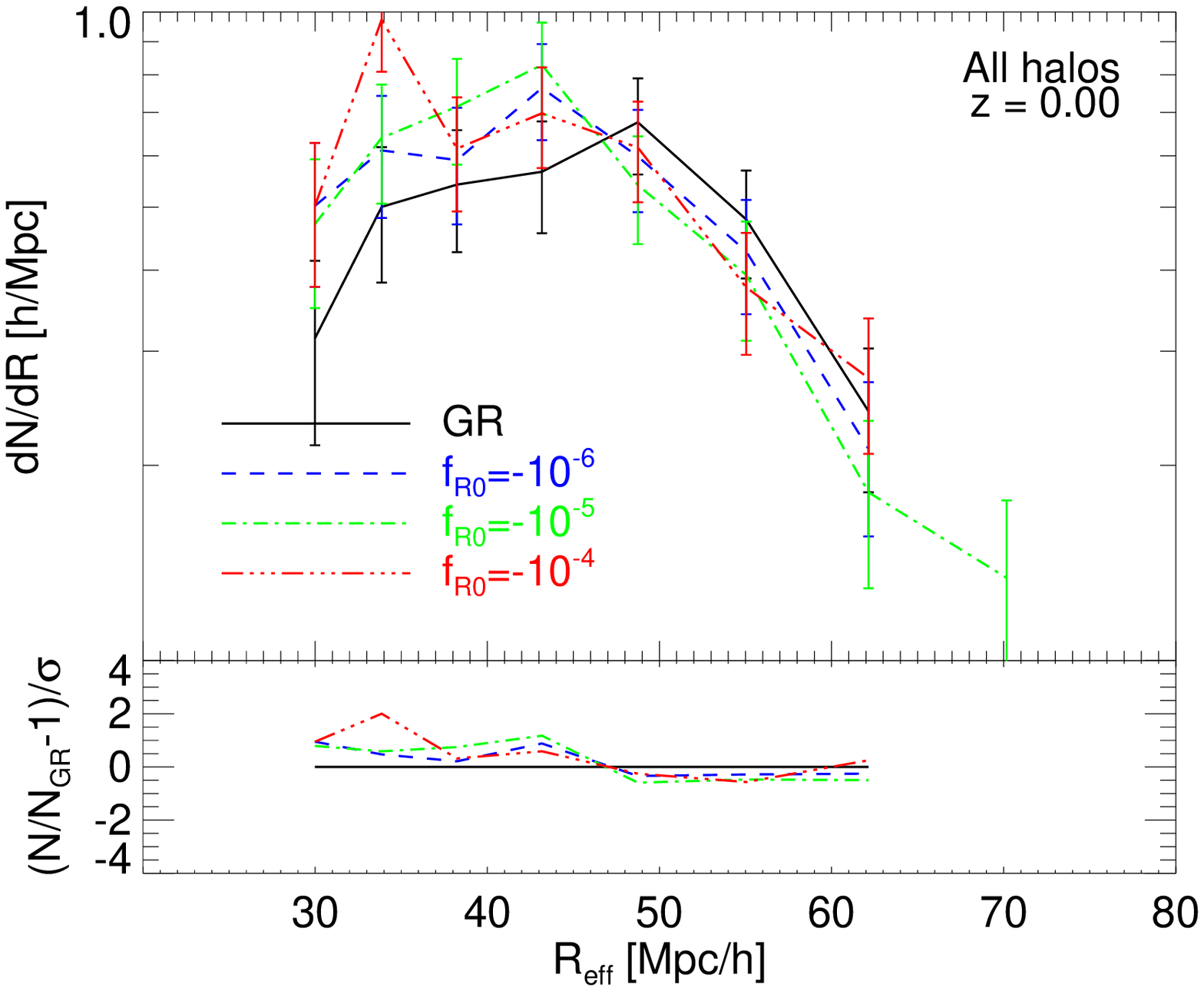}
\caption{{\em Upper panel} -- Void abundance detected in the CDM density field using VIDE and a random subsampling of the CDM particles with a density of $0.02$ particles per cubic $\rm Mpc/ \rm h$. {\em Lower Panel} -- The same quantity for the sample of voids identified in the FoF halo catalog.}\label{nvfig}
\label{nvfield}
\end{figure}

Due to the presence of the screening mechanism, cosmic voids can potentially be very powerful tools to constrain $f(R)$ models, as the central parts of the voids are expected to be unscreened due to their low matter density, while the void walls could be fully or partially screened. Hence voids provide the opportunity to test the transition between these two regimes. Several works have been focusing on measuring the void abundance and its deviation from the expectations of the standard $\Lambda$CDM model  in the context of various non-standard cosmologies (e.g. \citep{Li2011,Zivick2014,Cai2014,Sutter_etal_2015,Pollina_etal_2015}). In particular, for the case of $f(R)$ gravity, some of these works \citep[][]{Li2011,Zivick2014,Cai2014} have shown that there is an excess of large voids compared to $\Lambda$CDM, and that the excess is larger for larger values of the $f_{R0}$ parameter. This confirms the intuitive prediction that the action of the fifth force pushes the particles stronger towards the void wall \cite{Clampitt2013}, thereby more efficiently evacuating the void centres. However, most of these analysis are based on cosmic voids identified in some random subsampling of the dark matter density field, while voids are generally identified in galaxy surveys. The effect of the tracer bias and of the CDM particle sampling on analytical predictions for the void abundance is still an unsolved issue  for the standard $\Lambda$CDM case \citep[see e.g.][]{Jennings2013,ANP, Pollina_etal_2015,Nadathur_Hotchkiss_2015}. Recently, \cite{Cai2014} have compared  the abundance and the density profiles of voids identified using the halo catalogs of some $f(R)$ simulations to the case where voids are identified in a random subsample of CDM particles, finding that the deviations with respect to the $\Lambda $CDM cosmology are much weaker when halos are used as density tracers. A similar conclusion has been reached by \cite{Pollina_etal_2015} for interacting dark energy models.
In this paper, we focus on void abundances and profiles detected in the simulations of \cite{Puchweinetal} using the {\small VIDE} void finder \citep[][]{Sutter_etal_2014}, based on the {\small ZOBOV} code \cite{Neyrinck08}. Nonetheless, we adopt a slight modification \cite{Pollina_etal_2015} of the void hierarchy classification provided by {\small VIDE} which allows to eliminate possible pathological voids and to obtain a converging void abundance when changing the sparsity of the density tracers.

Specifically, for each $f(R)$ model and for the standard GR simulations we have run {\small VIDE} on both a random subsampling of the CDM particles (with an average density of $0.02$ particles per cubic Mpc$/h$), and on the positions of the FoF halos identified in the simulations (having an average density of about $8.4\times 10^{-4}$ halos per cubic Mpc$/h$), both at $z=0$. In order to avoid multiple counting of the volume included in sub-voids within a complex void hierarchy we have selected only the main voids, i.e. those voids that are not part of larger voids. Furthermore, among these voids we have defined two selection criteria to remove possible pathological voids (see \cite{Pollina_etal_2015}), namely requiring a minimum void density $\rho _{\rm v,min}$ lower than $0.2$ times the average density and a density contrast (i.e. the density ratio between the density minimum $\rho _{\rm v,min}$ and the void boundary $\rho _{\rm v,bound}$) larger than $1.57$ (see \cite{Neyrinck08}). If a given main void identified by {\small VIDE} does not fulfil both these criteria we remove it from our catalog and we promote all its first tier of sub-voids to the role of main voids. This selection procedure is then repeated until no pathological main voids are left in the sample. This procedure allows to obtain a robust sample of main voids whose abundance does converge (down to the resolution scale allowed by the density of the tracers) for different random sub-samplings of the original CDM density field or different densities of the FoF halo sample.

 In Fig.~\ref{nvfig} we show the size distribution of cosmic voids identified as a function of their effective radius $R_{\rm eff}$ defined from the void volume $V$ assuming spherical symmetry. The upper panel shows the result obtained for voids identified based on our random sub-sampling of the CDM particle distribution, while the lower panel displays the same quantity obtained from the FoF halo catalog. In each panel, the upper plot shows the number of voids as a function of their radius, where the error bars represent the Poisson error associated with the number counts in each bin, while the lower plot shows the significance of the relative difference between each $f(R)$ model and standard GR. It is interesting to notice that the two plots show a very different relative trend among the models. While in the halo voids sample we do not observe significant deviations between the different gravity models except for a mild excess of large voids for the $\Lambda$CDM and the $f_{R0}=-10^{-6}$ case, in the CDM voids sample the trend is the opposite and presents a much higher significance, with all the $f(R)$ scenarios showing a higher abundance of large voids as compared to $\Lambda $CDM. This stark difference in the relative behaviour of the expected size distribution of cosmic voids, when the latter are identified using a biased vs. an unbiased set of tracers of the underlying CDM density field, has been already pointed out by \cite{Cai2014} for the case of $f(R)$ gravity, and more recently by \cite{ANP, Pollina_etal_2015,Nadathur_Hotchkiss_2015}. This can be explained as a consequence of the different bias that a population of halos within the same mass range is found to have in different gravity models: while the additional gravitational force acting in $f(R)$ cosmologies is effective in evacuating the voids from the smooth CDM component, thereby increasing the amplitude of CDM density perturbations (and consequently the size and depth of cosmic voids, see the next section), the bias of collapsed objects is substantially reduced in these scenarios with respect to standard gravity, so that the net effect on amplitude and size of perturbations in the halos population is reduced (see\citep{Pollina_etal_2015}).
It is interesting to see that this effect is still present for different void finders. In \citep{Cai2014} the authors used a spherical shell to identify voids in the distribution of halos. In this case the authors had find a ratio of the number of voids in $f(R)$ gravity to GR of $N_{f(R)}/N_{\rm GR}\sim 0.3$ for $R\sim 45 {\rm Mpc}/h$ in the $f_{R0}=-10^{-4}$ case, while we do not measure a lower number of voids on this scale. Therefore using the abundance of voids to discriminate between GR and $f(R)$ gravity must be taken with some caution. Depending on the algorithm used to identify voids, the population of tracers (dark matter Vs. halos), and their number density, different results can be obtained.

\section{Void profiles}\label{secVP}

\begin{figure*}[ht]
\centering
\includegraphics[width=\textwidth]{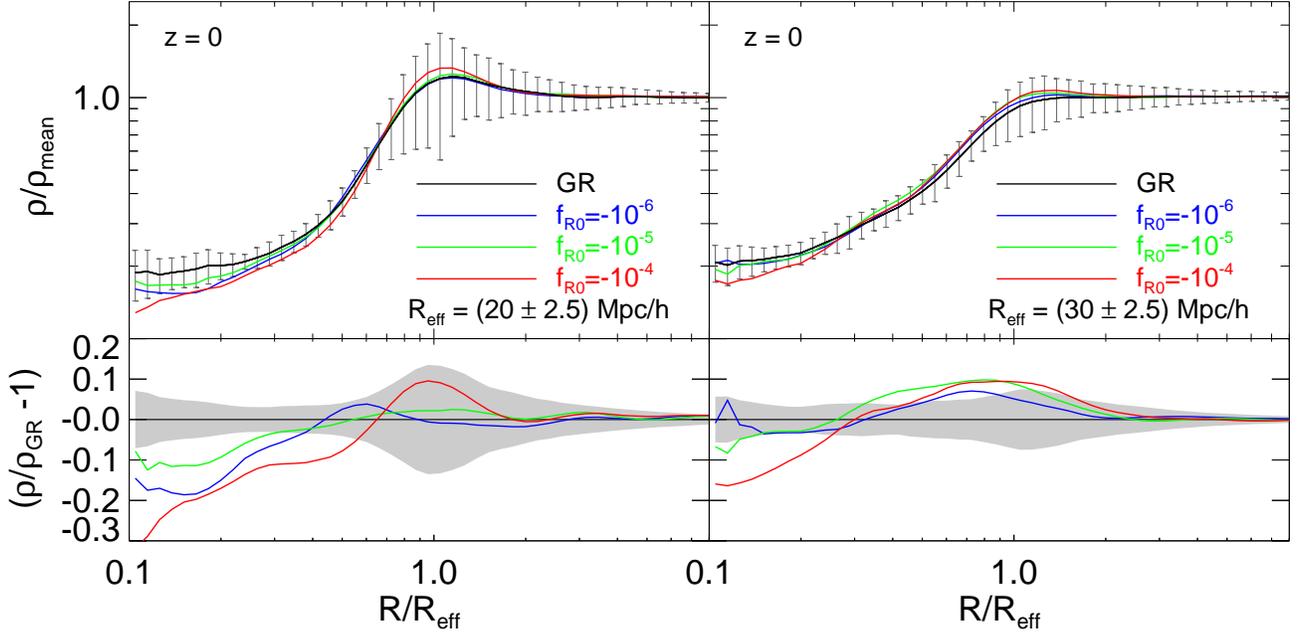}
\caption{Stacked density profiles using the void centres of each simulation.}
\label{void_profiles}
\end{figure*}

\begin{figure*}[ht]
\centering
\includegraphics[width=\textwidth]{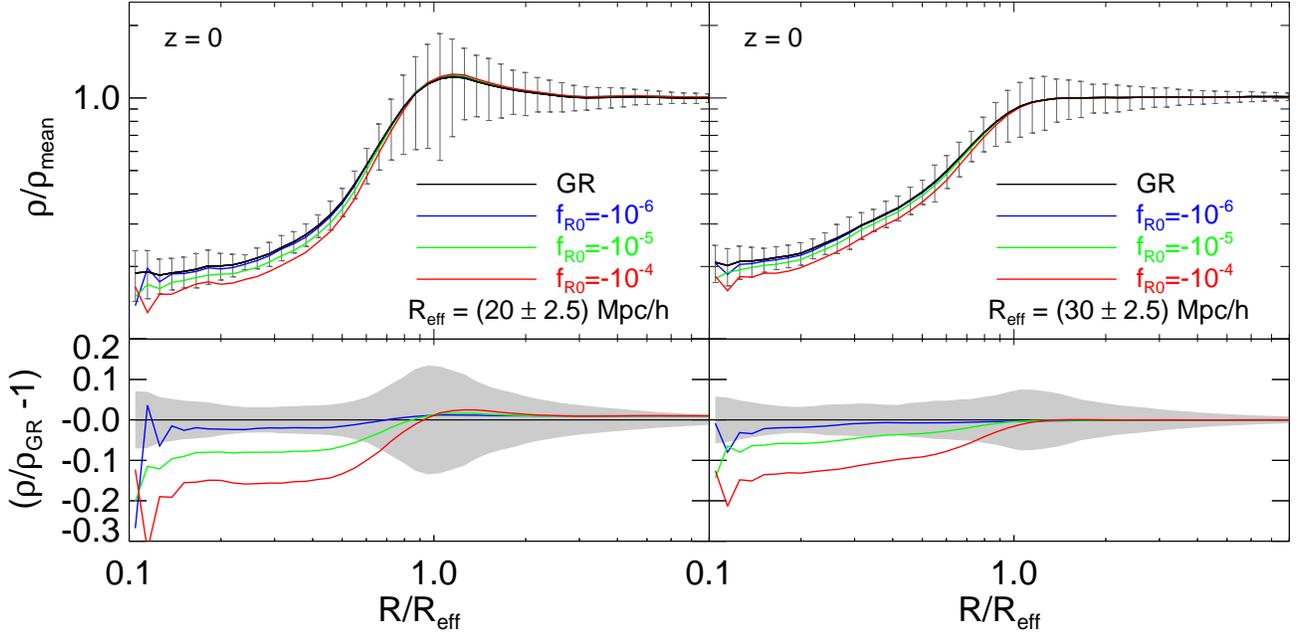}
\caption{Stacked density profiles using the void centres of the GR simulation.}
\label{void_profiles_GR}
\end{figure*}

Void profiles have been extensively studied in the recent literature. In particular, several studies (see e.g. \cite{ricciardelli2013, ricciardelli2014,Hamaus2014}) have claimed the self-similarity of the density profiles of cosmic voids in the standard $\Lambda $CDM model, which would make the shape of voids an ideal geometrical probe of the underlying cosmology. As mentioned before, in the context of $f(R)$ gravity, recent works (as e.g. \cite{Zivick2014,Cai2014,Zivick2015}) have shown that the extra force has the effect of evacuating faster the inner regions of cosmic voids, thereby producing steeper void profiles. A similar result is found for interacting dark energy models, even in the absence of a scale-dependent screening mechanism (\cite{Sutter_etal_2015,Pollina_etal_2015}). Here we compute the void density profiles in our set of $f(R)$ simulations using two different approaches: \\

(i) stacking density profiles using void centres of each simulation\\

(ii) stacking density profiles using the centres of the GR simulation.\\

 The latter approach allows us to to test more precisely the effect of $f(R)$ since all simulations start from the same initial conditions. Therefore, by doing this one-to-one comparison, we avoid fluctuations in the profiles due to the selection of voids using VIDE. Any deviations from the GR profiles are due to the $f(R)$ gravity.

 To this end, we first compute the spherically-averaged radial density
profile of each individual void by estimating the CDM density within a
series of logarithmically equispaced spherical shells centred in the
barycentre of each void and normalised to the void effective radius
$R_{\rm eff}$. We perform this procedure for two distinct ranges of void size, namely $R_{\rm eff} =\left\{ 20\,, 30\right\} \pm 2.5$ Mpc$/h$, considering 100 randomly selected voids in each case. Then we obtain the stacked profiles by averaging the overdensity in each radial bin over the 100 individual profiles.
The resulting profiles are displayed
in Fig.~\ref{void_profiles}, where we show the
comparison of the stacked density profiles obtained using the CDM void
catalog in standard GR and in the various realisations of $f(R)$ gravity. In the upper panels of Fig.~\ref{void_profiles} the error bars represent the corrected 
sample standard deviation computed on the 100 randomly selected voids, while in the lower panels we show the relative difference between the models and the grey-shaded area represents the $2 \sigma$ confidence on the relative difference based on the standard deviation of the mean density profiles computed through a {\em bootstrap} resampling technique with $1000$ re-samples of the $100$ individual profiles. As the figure shows, $f(R)$ voids present a systematically lower density in the inner part of the void with a significant deviation from the GR behavior.

In Fig.~\ref{void_profiles_GR} we show the same results obtained by computing the individual spherically-averaged profiles, always assuming, i.e. for all gravity models, the void centre computed from the GR simulation. This results in a clearer hierarchy among the models in the overdensity deviation from GR that appears in the central part of the voids, with the deviation increasing for increasing values of the $f_{\rm R0}$ parameter. As one can see from the figure, only the $f_{\rm R0}=-10^{-6}$ model is found to be consistent at $2 \sigma$ with the GR result, while the other models are both clearly distinguishable from the standard case.

To conclude this section, the imprint of $f(R)$ in the void profiles is enhanced near the centres of voids and for smaller void sizes with significant deviations from the GR case except for the weakest modification of gravity considered in the present work.

\section{Conclusions}\label{secConclu}

(i) In this paper we performed an analysis of the $f(R)$ imprints on the large scale structure of the Universe in the non-linear domain. We started by looking at the halo abundance measured in N-body simulations. We used the theoretical model of \cite{Koppetal} to fit the multiplicity function and found a good accuracy with the N-body simulations up to $z=1.5$. Overall we found a larger number of massive halos in $f(R)$ gravity, which agrees with previous studies (e.g. \citep{LLKZ}\cite{Cai2014}. However, unlike previous studies (e.g. \citep{LLKZ}) we found the drifting diffusive barrier mass function reproduces the $f(R)$ N-body simulations also in the low halo masses regime. 
\medskip

(ii) We extended the theoretical multiplicity function to the conditional one. We used Monte Carlo random walks to ensure the robustness of our prediction and we found a very good correspondence for different large scale perturbations at the $\sim 5\%$ level. From this prediction we derived the linear halo bias  for $f(R)$ gravity. The latter is fixed by the parameters of the halo mass function and in particular the linear criteria of collapse, (stochastic barrier). Because most of the stochastic nature of the barrier is due to the excursion set framework \cite{ARSC}\cite{AWWR}, our prediction of the conditional mass function can be easily extended to any theory of gravity, providing the linear spherical criteria for halos to collapse.
\medskip

(iii) We measured the halo density profiles in the $f(R)$ and GR N-body simulations. We found a deviations from GR depending on the halo masses, redshift and $f_{\rm R0}$ parameters. This deviation tends to decrease for massive halos at low redshift (e.g. $2 \times 10^{14} M_\odot /h$ at $z=0$). However, it is unclear how to model the trend of these profiles considering that the amount of the deviations are not necessarily proportional to the modified gravity parameter $f_{R0}$. In \cite{Balmes}, the author study how the NFW profile varies with cosmology and quantify this information in an observable quantity named sparicity. It would be interesting to see if this quantity can be used to probe the $f(R)$ gravity. 

\medskip

(iv) Neglecting the difference between GR and $f(R)$ halo profiles at fixed halo mass, we used the halo model with our predictions of the linear halo bias and halo mass function to predict non-linear clustering. For the GR case we found an agreement on the order of $\sim 10 \%$ at $z=0$ and $20\%$ at $z=1.5$ for $k\leqslant 2 h{\rm Mpc}^{-1}$. On the other hand for the $f_{\rm R0}=-10^{-4}$ case the agreement between our prediction and the N-body measurement is $\sim 10\%$ at $z=0$ and only $\sim 35 \%$ at $z=1.5$. Overall we provide a simple prediction of the non-linear power spectrum which qualitative agree with the $f(R)$ simulation at $z=0$. 
\medskip

(v) We studied the void number density detected in $f(R)$ and GR simulations. We tested the effect of the bias and sampling by considering as tracers both dark matter particles and halos. We found more voids in the dark matter density field in the case of $f(R)$ gravity. This was already observed in previous studies and with different void finders (e.g.\citep{Cai2014}). However, this effect is erased when we identified voids using halos. In this case the ratio between $f(R)$ voids to GR depends on the number density and the void finder.  
\medskip

(vi) We tested the imprint of $f(R)$ on the void density profiles using a standard stacking technique and a one-to-one comparison between the GR profiles and the $f(R)$ cases. In all case we found that void densities are lower in $f(R)$ compared to GR. This qualitatively agree with \citep{Cai2014}\citep{Zivick2014}\citep{Zivick2015}.

\section*{Acknowledgments}
We acknowledge discussions with Katharina Wollenberg at the initial stages of this work.
Part of this research was conducted by the Australian Research Council
Centre of Excellence for All-sky Astrophysics (CAASTRO), through
project number CE110001020.
 I. Achitouv and J. Weller also acknowledges
support from the Trans-Regional Collaborative Research
Center TRR 33 ``The Dark Universe'' of the
Deutsche Forschungsgemeinschaft (DFG). M Baldi acknowledges support from the Italian Ministry for Education, University and Research
(MIUR) through the SIR individual grant SIMCODE, project number RBSI14P4IH. E. Puchwein is grateful for support by the Kavli Foundation and the ERC Advanced Grant 320596 ‘The Emergence of Structure during the epoch of Reionization

\end{document}